\documentclass[11pt,a4paper]{article}
\pdfoutput=1
\usepackage{jcappub}

\usepackage{epsfig,amssymb,bm, graphicx, amsmath}
\usepackage[normalem]{ulem}

\usepackage{color}

\usepackage{enumerate,enumitem}

\def\o{^{\text{obs}}}

\newcommand{\PRE}[1]{}
\newcommand{\Expect}[1]{\left\langle #1 \right\rangle}

\newcommand{\beas}{\begin{eqnarray*}}
\newcommand{\eeas}{\end{eqnarray*}}

\newcommand{\nn}{\nonumber}
\newcommand{\be}{\begin{equation}}
\newcommand{\ee}{\end{equation}}
\newcommand{\bea}{\begin{eqnarray}}
\newcommand{\eea}{\end{eqnarray}}

\definecolor{darklightsabergreen}{rgb}{0.0, .49, 0.06}

\title{Cosmic Variance of the Spectral Index from Mode Coupling}
\author[a]{Joseph Bramante,}
\author[a]{Jason Kumar,}
\author[b]{Elliot Nelson,}
\author[b,c]{Sarah Shandera}
\affiliation[a]{Department of Physics and Astronomy, University of Hawaii, Honolulu, HI 96822, USA}
\affiliation[b]{Institute for Gravitation and the Cosmos, The Pennsylvania State University, University Park, PA 16802, USA}
\affiliation[c]{Kavli Institute for Theoretical Physics, University of California Santa Barbara,
Santa Barbara, CA 93106, USA}

\emailAdd{bramante@hawaii.edu, jkumar@hawaii.edu, eln121@psu.edu, shandera@gravity.psu.edu}
\abstract{
We demonstrate that local, scale-dependent non-Gaussianity can generate cosmic variance uncertainty in the observed spectral index of primordial curvature perturbations. In a universe much larger than our current Hubble volume, locally unobservable long wavelength modes can induce a scale-dependence in the power spectrum of typical subvolumes, so that the observed spectral index varies at a cosmologically significant level ($|\Delta n_s|\sim\mathcal{O}(0.04)$). Similarly, we show that the observed bispectrum can have an induced scale dependence that varies about the global shape. If tensor modes are coupled to long wavelength modes of a second field, the locally observed tensor power and spectral index can also vary. All of these effects, which can be introduced in models where the observed non-Gaussianity is consistent with bounds from the {\it Planck} satellite, loosen the constraints that observations place on the parameters of theories of inflation with mode coupling. We suggest observational constraints that future measurements could aim for to close this window of cosmic variance uncertainty.
}

\preprint{UH511-1209-2013, IGC-13/7-3, CETUP2013-005, NSF-KITP-13-235}

\begin{document}
\maketitle

\section{Introduction}
\label{sec:intro}
The temperature fluctuations of the Cosmic Microwave Background (CMB) have been measured to a remarkable precision by the {\it Planck} satellite \cite{Ade:2013ktc, Ade:2013uln, Ade:2013tta}. Two of the inferred properties of the primordial scalar curvature fluctuations have particularly important implications for theories of the very early universe: the strong evidence for a red tilt in the primordial power spectrum and the limits on the amplitude of any primordial non-Gaussianity. This evidence from the power spectrum and the bispectrum supports the simplest models of inflation with a single degree of freedom and no significant interactions, but does not yet rule out other possibilities. Future constraints or measurements of non-Gaussianity will continue to provide significant avenues to differentiate models of inflation. Interestingly, while non-Gaussian signatures offer a way to distinguish inflation models with identical power spectra, mode coupling also introduces a new and significant uncertainty in matching observations to theory \cite{Linde:2005yw,Demozzi:2010aj, Nelson:2012sb, Nurmi:2013xv, LoVerde:2013xka}. In a universe much larger than our current Hubble scale, our local background may not agree with the global background used to define homogeneous and isotropic perturbations on a much larger region generated from inflation. If modes are coupled, the observed properties of the statistics in our Hubble volume will depend on the long wavelength background which is not independently observable to us. That is, our local statistics may be biased.

This cosmic variance due to mode coupling was discussed for curvaton models in \cite{Linde:2005yw,Demozzi:2010aj} and was recently explored more generally in \cite{Nelson:2012sb, Nurmi:2013xv, LoVerde:2013xka,Byrnes:2013qjy} for non-Gaussianity generated by arbitrary non-linear but local transformations of a Gaussian field. This non-Gaussian family is described by the local ansatz \cite{Salopek:1990jq,Komatsu:2001rj} for the curvature perturbation $\zeta$:
\be
\zeta(\mathbf{x})=\zeta_G(\mathbf{x})+\frac{3}{5}f_{\rm NL}[\zeta_G(\mathbf{x})^2-\langle \zeta_G(\mathbf{x})^2 \rangle] +\frac{9}{25}g_{\rm NL}\zeta_G(\mathbf{x})^3+\dots,
\label{local}
\ee
where $\zeta_G(\mathbf{x})$ is Gaussian and $f_{\rm NL}$, $g_{\rm NL}$, etc. are constants. For curvature perturbations of this type the amplitude of fluctuations and the amplitude of non-Gaussianity (the observed $f_{\rm NL}$, $g_{\rm NL}$, $\dots$) vary significantly throughout the entire inflationary region. This is true even with globally small fluctuations, weak non-Gaussianity, and of order 10-100 extra e-folds of inflation.

In this paper we explore further implications of mode coupling in the primordial fluctuations. We focus on a generalization of the local ansatz above, allowing $f_{\rm NL}$, $g_{\rm NL}$, etc. to be scale-dependent. In that case, the curvature fluctuations measured in subvolumes do not all have the same spectral index and bispectral indices as the parent theory. In other words, the possibility of mode-coupling, even at a level consistent with {\it Planck} bounds on non-Gaussianity, relaxes the restrictions that the precisely measured red tilt places on the theory of the primordial fluctuations.

This paper and \cite{Nelson:2012sb, Nurmi:2013xv, LoVerde:2013xka} work out the observational consequences of mode coupling in the post-inflation curvature fluctuations without asking which dynamics generated the fluctuations. We work with curvature perturbations that are assumed to be output by some inflationary model, and the mode-coupling effects we discuss are only significant if modes of sufficiently different wavelengths are physically coupled. Purely single field models of inflation do not generate such a coupling \cite{Maldacena:2002vr,Creminelli:2004yq,Pajer:2013ana}, and in Section \ref{sec:NonLocalBispect} we demonstrate how to see directly from the shape of the bispectrum that single field type bispectra do not lead to cosmic variance from subsampling. So, although the curvature perturbations in Eq.(\ref{local}) have a single source, the inflationary scenario they come from must be multi-field. For example, the distribution of locally observed non-Gaussian parameters $f_{\rm NL}$ in the curvaton scenario was studied early on in \cite{Linde:2005yw,Demozzi:2010aj}. From the point of view of using observations to constrain inflationary scenarios that generate local non-Gaussianity, the variation allowed in local statistics means that the parameters in an inflaton/curvaton Lagrangian with local type mode coupling are not exactly fixed by observations. Given a Lagrangian and a restriction on the maximum size of a post-inflation region with small fluctuations, there is a probability for a set of observations (e.g., characterized by the power spectrum and $f_{\rm NL}$, $g_{\rm NL}$, etc.) in a patch of the universe the size we see today. Put the other way around, the parameters in the `correct' Lagrangian need only fall within the range that is sufficiently likely to generate a patch with the properties we observe. Although the {\it Planck} bounds on local type non-Gaussianity are quite restrictive, they are not restrictive enough to eliminate the possibility of this effect: the data are also consistent with an application of the cosmological principle to a wider range of non-Gaussian scenarios with more than the minimum number of e-folds.

In the rest of this section, we review the role of unobservable infrared modes coupled to observables in cosmology and particle physics. We also set up our notation and briefly review previous results for the local ansatz. In Section \ref{sec:Generalizing} we introduce a scale-dependent two-source ansatz, which changes the momentum dependence of the correlation functions. We compute features of the power spectrum and bispectrum observed in subvolumes and show that the locally observed spectral index of primordial scalar perturbations ($n_s^{\rm obs}$) can be shifted by scale dependent coupling to modes that are observationally inaccessible. In Section \ref{sec:ObsConsequences} we illustrate how cosmic variance from mode-coupling affects the relationship between observation and theory for the spectral index and the amplitude of the power spectrum and bispectrum. The reader interested only in the consequences and not the detailed derivations can skip to those results. There we illustrate, for example, that a spectrum which is scale-invariant on observable scales may look locally red or blue, and a red spectrum may look locally redder, scale-invariant, or blue when scale-dependent non-Gaussianity is present. It is unlikely to find subvolumes with an observed red tilt inside of a large volume with nearly Gaussian fluctuations with a blue power spectrum. However, a large volume with a blue power spectrum on observable scales due to a significant non-Gaussian contribution may have subvolumes with power spectra that are nearly Gaussian and red. We similarly show that the scale dependence of the bispectrum and higher order spectra in our Hubble volume can be shifted by non-Gaussian correlations with modes that are observationally inaccessible.

Section \ref{sec:NonLocalBispect} calculates the effect of a generic factorizable bispectrum on the amplitude and scale dependence of the power spectrum in subvolumes and verifies that not all bispectra lead to a variation in the locally observed statistics. The effects of mode coupling on the power spectrum and spectral index of tensor modes is considered in Section \ref{sec:tensors}. We summarize our results in Section \ref{sec:conclude} and suggest future observational limits that could rule out the need to consider these statistical uncertainties in using observations to constrain (the slow-roll part of) inflation theory.

\subsection{Long wavelength modes in cosmology and particle physics}
There has been a great deal of recent literature on mode coupling in the primordial fluctuations, and an infrared scale appearing in loop corrections. In some cases, there may appear to be a naive infrared divergence as this scale is taken to be infinitely large. However, in calculating quantities observable within our own universe, such divergences clearly cannot be physical. For an example of a treatment of the infrared scale from an astrophysical perspective, see \cite{McDonald:2008sc}. See \cite{LoVerde:2011iz} for a treatment in the simulation literature. Previous discussions in the context of single field inflation, but with a flavor similar to our work here, can be found in \cite{Gerstenlauer:2011ti, Giddings:2011zd}.

The meaning of the infrared scale appears in computing $n$-point functions \textit{averaged locally in a given Hubble volume}. Any long-wavelength modes come outside the expectation value and contribute a constant depending on the particular realization of the long wavelength modes in the local patch. These local $n$-point functions may differ from the global $n$-point functions due to the influence of long-wavelength background fluctuations, and the precise relationship between the local and global statistics depends on the local background. Of course, if one averages over all the individual subvolumes, the statistics must recover those initially defined in the large volume regardless of the scale chosen for the subvolume \cite{Lyth:2007jh}.

The relevance of unmeasurable infrared modes is well studied in non-cosmological contexts. A well known example in particle physics is the appearance of Sudakov log factors in the cross section for electron scattering in quantum electrodynamics.  One finds infra-red divergences in both the one-loop
correction to the cross-section for exclusive $e^- e^- \rightarrow e^- e^-$ scattering, and in the tree-level cross-section
for the process $e^- e^- \rightarrow e^- e^- \gamma$ where electrons scatter while emitting a photon.  In both cases, the
divergence is associated with very long wavelength modes of the electromagnetic field, and can be regulated by
introducing an infra-red cutoff (analogous to the infrared scale that appears in cosmological calculations).  But neither of the scattering processes
is observable in and of itself, because one cannot distinguish events in which a photon is emitted from
events in which it is not if the wavelength of the photon is much larger than the size of the detector. Instead, the
physically measurable quantity is the cross section for electrons to scatter while emitting no photons with energy larger than
the energy resolution of the detector, $E_{res}$.  For this quantity, the infra-red divergences (and dependence on the infra-red
cutoff) cancel, but a logarithmic dependence on $E_{res}$ is introduced (the Sudakov log factor).
This dependence is physically meaningful and represents the fact that,
if the energy resolution of the detector is degraded, then events in which a soft photon is emitted may appear to be exclusive $e^- e^- \rightarrow
e^- e^-$ scatters because the long-wavelength photon can no longer be resolved by the detector.  In the context
of primordial curvature fluctuations, the energy resolution is equivalent to the scale of the subvolume and is ultimately limited by the size of the
observable universe; a mode with a wavelength much
longer than the observable universe cannot be distinguished from a zero-mode. Furthermore, we only have one universe - we are stuck with one particular set of unmeasurable long wavelength modes. Predictions for observable consequences of inflation models with mode-coupling should account for the fact that observations cannot access information about larger scales.

In this work, we are assuming that we have been given a set of non-Gaussian inhomogeneities as output from a dynamical model for generating them. We limit ourselves to considering curvature perturbations on a spatial slice, defined with a notion of time appropriate for observations made in our universe after reheating, on which there are small perturbations on a homogeneous and isotropic background. We suppose that the spatial volume over which this description holds is unknown, but that it may be much larger than what we can currently observe. However, we do not consider a spatial slice that is significantly inhomogeneous on large scales so our results should be adjusted to apply to scenarios that enter the eternal inflation regime. Although our calculations contain integrals over momenta, there are no corresponding time integrals. Our momenta integrals are not the `loops' from dynamics, but merely add up the effects of all the modes coupled to a mode of interest at a particular time. Our analysis will focus on how the fact that long wavelength modes are unobservable affects our ability to compare a particular set of local observations to a model prediction for the larger, statistically homogeneous slice.

\subsection{Statistics of $\zeta$ in a subsample volume with local type non-Gaussianity}
\label{sec:LSsplit}
The curvature perturbation in either a large volume ($L$) or a subsample volume ($M$)
is defined as the fractional fluctuation in the scale factor $a$,
\bea
a(x)&=&\langle a \rangle_L(1+\zeta(x))\;,\;\;\;x\in \text{Vol}_L  \label{fracfluctuation}\\
&=&\langle a \rangle_M(1+\zeta^{\text{obs}}(x))\;,\;\;\;x\in \text{Vol}_M,
\eea
where $\langle \, \rangle_{L,M}$ refers to the value of a field averaged over the volume
$L$ or $M$, respectively.
We assume $|\zeta|,|\zeta^{\text{obs}}|\ll1$, and thus keep only the linear term.
Throughout the paper, we will denote with an ``obs'' superscript quantities as defined within a subsample volume such as the observable universe,
which do not correspond (except perhaps by a coincidence of values) to quantities in the larger volume.
Since $\langle a \rangle_M=\langle a \rangle_L(1+\langle\zeta\rangle_M)$, we see that $\zeta$ and $\zeta^{\text{obs}}$ are related by
\be
1+\zeta(x)=(1+\langle\zeta\rangle_M)(1+\zeta^{\text{obs}}(x))\;,\;\;\; x\in \text{Vol}_M. \label{zeta}
\ee
Dividing $\zeta$ into long and short-wavelength parts compared to the scale $M$, $\zeta\equiv\zeta_l+\zeta_s$, and considering one particular subvolume we have \cite{Nelson:2012sb}
\be
\zeta^{\text{obs}}=\zeta_s/(1+\zeta_{l})\;,\;\;\; x\in \text{Vol}_M, \label{zetaLs,M}\;.
\ee
We have replaced the mean value $\langle\zeta\rangle_M$ with the field smoothed on scale $M$, $\zeta_{l}$ (the only difference being the real space vs.~Fourier
space top-hat window functions). $\zeta_l$ takes a particular constant value for the subsample in question. For the remainder of the paper, averages $\langle \ \rangle$ are taken over the large volume $L$, and averages over the small volume $M$ are represented by a subscript ``$l$".

In either volume, the two-point function defines the power spectrum,
\be
\langle \zeta_{\mathbf{k}_1} \zeta_{\mathbf{k}_2}\rangle\equiv(2\pi)^3\delta^3(\mathbf{k}_1+\mathbf{k}_2)P_{\zeta}(k)\;.
\ee
We will consider homogeneous and isotropic correlations, so the bispectrum ($B_{\zeta}(k_1,k_2,k_3)$) is defined by
\bea
\Expect{\zeta_{\mathbf{k}_1} \zeta_{\mathbf{k}_2} \zeta_{\mathbf{k}_3}} &\equiv & (2 \pi)^3 \delta^3(\mathbf{k}_1 + \mathbf{k}_2 + \mathbf{k}_3)B_{\zeta}(k_1,k_2,k_3)\;.
\eea
From Eq.~\eqref{zetaLs,M}, we see that the power spectra in the two volumes are related by
\be
P^{\text{obs}}_{\zeta}(k)=P_{\zeta}(k)/(1+\zeta_{l})^2.
\label{rP}
\ee
The amplitude of linearized fluctuations is thus rescaled by a factor of $1+\zeta_{l}$ due to the shift in the local background by the same factor: fluctuations appear smaller in overdense regions and larger in underdense regions. In general, an $n$-point function averaged in the small volume differs from the corresponding $n$-point function averaged in the large volume by a factor $(1+\zeta_{l})^{-n}$. However, this shift does not affect the level of non-Gaussianity in the small volume, as quantified for example by the dimensionless connected moments $\mathcal{M}_n\equiv\langle\zeta(\mathbf{x})^n\rangle_c/\langle\zeta(\mathbf{x})^2\rangle^{n/2}$, nor does it affect the shapes of the $n$-point functions, which will be our focus, but only reflects the rescaling of $\zeta$. In what follows, we will therefore drop factors of $1+\zeta_{l}$ in expressions for spectral indices, and also in expressions for $n$-point functions, to which these factors yield corrections smaller than our level of approximation.

\begin{figure}
\caption{The size of background fluctuations $\langle\zeta_{Gl}^2\rangle^{1/2}$ which bias local statistics, as a function of the number of superhorizon e-folds. Red and blue curves assume red ($n_{\zeta}=0.96$) and blue ($n_{\zeta}=1.02$) spectral indices, while black curves show the scale-invariant case. Solid curves fix $\mathcal{P}_G(M^{-1})=\mathcal{P}\o_{\zeta}\simeq2.7\times10^{-9}$, while dashed curves fix $\mathcal{P}_G(M^{-1})=\frac{1}{10}\mathcal{P}\o_{\zeta}$, as in the case where a second source contributes the dominant fraction of the fluctuations.}
\centering
\includegraphics[width=.7\textwidth]{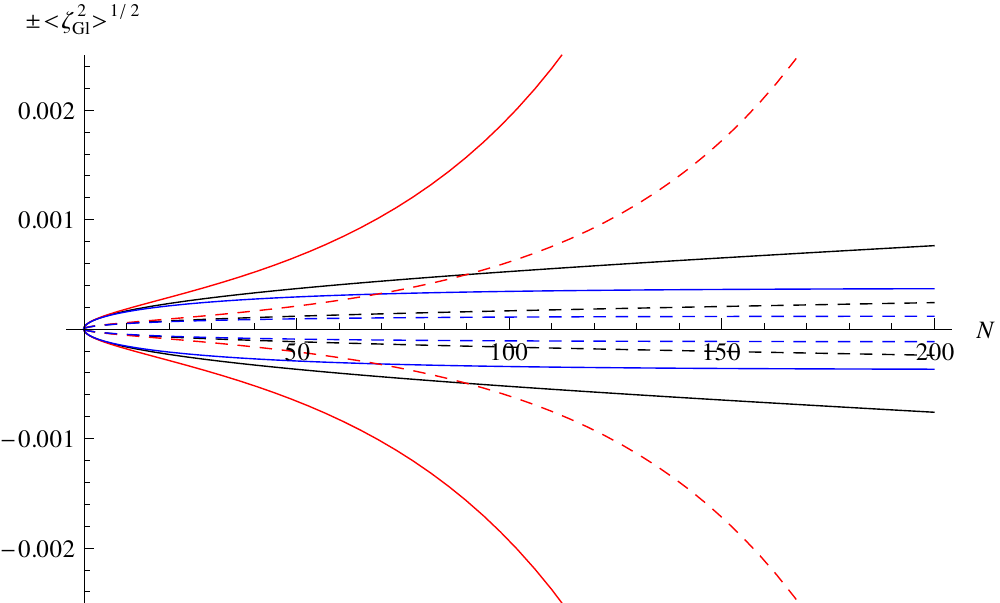}
\label{Bias}
\end{figure}

Suppose the curvature perturbation in the large volume is given by the local ansatz\footnote{This definition for $g_{\rm NL}$  differs slightly from that in \cite{LoVerde:2013xka}. The difference is irrelevant for our purposes here and this choice simplifies the notation somewhat.}
\be
\zeta(\mathbf{x})=\zeta_G(\mathbf{x})+\frac{3}{5}f_{\rm NL}[\zeta_G(\mathbf{x})^2-\langle \zeta_G(\mathbf{x})^2 \rangle] +\frac{9}{25}g_{\rm NL}\zeta_G(\mathbf{x})^3+\dots,
\label{eq:localfamily}
\ee
where $\zeta_G$ is a Gaussian field.
Splitting the Gaussian field into long- and short-wavelength modes in comparison to the scale of the subvolume, $\zeta_G=\zeta_{Gl}+\zeta_{Gs}$, the long-wavelength pieces of higher order terms can be recollected in the coefficients of lower order terms. The curvature perturbation observed in a subvolume, $\zeta^{\text{obs}}$, is
\be
\zeta^{\text{obs}}(\mathbf{x})=\zeta_G^{\text{obs}}(\mathbf{x})+\frac{3}{5}f_{\rm NL}^{\text{obs}}[\zeta_G^{\text{obs}}(\mathbf{x})^2-\langle \zeta_G^{\text{obs}}(\mathbf{x})^2 \rangle] + \frac{9}{25}g_{\rm NL}^{\text{obs}}\zeta_G^{\text{obs}}(\mathbf{x})^3+\dots,
\ee
The coefficients $f_{\rm NL}^{\text{obs}}$, $g_{\rm NL}^{\text{obs}}$, etc., and the power spectrum now depend on the particular realization of long-wavelength modes for the subvolume \cite{LoVerde:2013xka}:
\bea
P\o_{\zeta}(k)&=&\left[1+\frac{12}{5}f_{\rm NL}\langle\zeta_G^2\rangle^{1/2}B+\mathcal{O}(f_{\rm NL}^2\langle\zeta_G^2\rangle)\right]P_{G}(k), \label{Pobs} \\
f_{\rm NL}^{\text{obs}}&=& f_{\rm NL}+\frac{9}{5}g_{\rm NL}\langle\zeta_G^2\rangle^{1/2}B-\frac{12}{5}f_{\rm NL}^2\langle\zeta_G^2\rangle^{1/2}B+\mathcal{O}(f_{\rm NL}^3\langle\zeta_G^2\rangle), \label{fNLobs}
\eea
where we have defined the power spectrum $P_G$ of the Gaussian field $\zeta_G$,
\be
\langle \zeta_{G,\mathbf{k}_1} \zeta_{G,\mathbf{k}_2}\rangle\equiv(2\pi)^3\delta^3(\mathbf{k}_1+\mathbf{k}_2)P_{G}(k)\;,
\ee
and the bias for a given subvolume (in a fixed size large volume) as
\be
B\equiv\frac{\zeta_{Gl}}{\langle\zeta_{G}^2\rangle^{1/2}}. \label{bias}
\ee
The bias is larger for more rare fluctuations and increases as the size of the subvolume considered is decreased. Leading contributions from superhorizon modes in Eqs. \eqref{Pobs}, \eqref{fNLobs} then go like $f_{\rm NL}\langle\zeta_G^2\rangle^{1/2}B$, where $f_{\rm NL}\langle\zeta_G^2\rangle^{1/2}$ is the level of non-Gaussianity in the large volume. As the size of the subsample approaches the smallest measurable scale (which means there are no measurable modes within the subsample), the bias for average volumes asymptotes to 1. To compare different large volume theories, with different spectral indices and different sizes for the large volume, notice that the degree of bias in any subvolume is also sensitive to the IR behavior of the power spectrum since
\bea
\langle\zeta_{Gl}^2\rangle=\mathcal{P}_{G}(M^{-1})\frac{1-e^{-(n_{\zeta}-1)N}}{n_{\zeta}-1},
\label{zetaGl^2}
\eea
where $N=\ln(L/M)$ is the number of superhorizon e-folds, $\mathcal{P}_{G}(k)\equiv (k^3 / 2\pi^2)P_{G}(k)$ is the dimensionless power spectrum, and we have assumed a constant spectral index $n_{\zeta}\equiv\frac{d\ln\mathcal{P}_G}{d\ln k}$ in evaluating the integral. This is the running if the \textit{Gaussian} field $\zeta_G$, which is distinguished from the \textit{total} running $n_s\equiv\frac{d\ln\mathcal{P}_{\zeta}}{d\ln k}$. For a red tilt, $n_{\zeta}<1$, with enough superhorizon modes, $N\gtrsim|n_{\zeta}-1|^{-1}$, the cumulative power of long-wavelength modes makes the degree of bias much greater than in the scale-invariant case \cite{LoVerde:2013xka}. The relationship between the parameters describing the fluctuations (e.g., $\mathcal{P}$, $f_{\rm NL}$) measured in a single small volume and those in the large volume depends in an unobservable way on the unknown IR behavior of the power spectrum. In Figure~\ref{Bias} we show the dependence of $\langle\zeta_{Gl}^2\rangle$ on the number of superhorizon e-folds for different values of the spectral index.

As detailed in \cite{Nelson:2012sb, Nurmi:2013xv, LoVerde:2013xka}, the coupling of long and short wavelength modes always present in an arbitrary member of the family in Eq.(\ref{eq:localfamily}) (that is, arbitrary values of the coefficients) means that small sub-volumes tend to look like the local ansatz with `natural' coefficients, $f_{\rm NL}^{\text{obs}}(\mathcal{P}_{\zeta}^{\text{obs}})^{1/2}\ll1$ and higher order terms falling off by the same small ratio. This effect was illustrated several years ago for the case of a purely quadratic term in \cite{Boubekeur:2005fj}. In addition, although the shapes of the correlation functions from arbitrary members of the local family are not identical they all have a sizable amplitude in squeezed configurations (e.g, $k_1\ll k_2\sim k_3$ for the bispectrum).

\section{Subsampling the local ansatz with scale-dependence in single- and multi-source scenarios}
\label{sec:Generalizing}
\subsection{The power spectrum} \label{sec:powerspectrum}
Next, we consider mode coupling effects for a generalized local ansatz for the real space curvature that has multiple sources with scale-dependent coefficients:
\bea
\label{eq:GenLocalRealSpace}
\zeta(\mathbf{x})&=&\phi_G(\mathbf{x})+ \sigma_G(\mathbf{x})+\frac{3}{5}f_{\rm NL}\star[\sigma_G(\mathbf{x})^2-\langle\sigma_G(\mathbf{x})^2\rangle]+\frac{9}{25}g_{\rm NL}\star{\sigma}_G(\mathbf{x})^3+...
\eea
where the dots also contain terms that ensure $\langle\zeta(\mathbf{x})\rangle=0$. We have defined the fields $\phi_G$ and $\sigma_G$ to absorb any coefficient of the linear terms, which typically appear, for example, in relating the inflaton fluctuations to the curvature. The Fourier space field, which we will use to do the long and short wavelength split, is
\bea
\zeta(\mathbf{k}) &=& \phi_G(\mathbf{k})+\sigma_{G}(\mathbf{k})+\frac{3}{5}f_{\rm NL}(k)\int \frac{d^3p}{(2\pi)^3} \sigma_{G}(\mathbf{p})\sigma_{G}(\mathbf{k}-\mathbf{p})
\nonumber \\
&+& \frac{9}{25} g_{\rm NL}(k)\int \frac{d^3p_1}{(2\pi)^3}\frac{d^3p_2}{(2\pi)^3} \sigma_{G}(\mathbf{p}_1)\sigma_{G}(\mathbf{p}_2)\sigma_{G}(\mathbf{k}-\mathbf{p}_1-\mathbf{p}_2)
+\cdots
\label{zetak}
\eea
For most of the paper, we will take $f_{\rm NL}$, $g_{\rm NL}$, etc. to be weakly scale-dependent functions:
\bea
\label{eq:nfng}
f_{\rm NL}(k) &=& f_{\rm NL}(k_p)\left(\frac{k}{k_p}\right)^{n_f}\\\nonumber
g_{\rm NL}(k) &=& g_{\rm NL}(k_p)\left(\frac{k}{k_p}\right)^{n_g}\;.
\eea

To determine the mapping between statistics in subsamples to those in the large volume, we follow the same procedure as in Section \ref{sec:LSsplit}, splitting the Gaussian part of the curvature perturbation into long and short wavemode parts: $\phi_G \equiv \phi_{Gl} + \phi_{Gs} $, $\sigma_{G} \equiv \sigma_{Gl} + \sigma_{Gs} $. The division happens at the intermediate scale $M$, which we take to be roughly the largest sub-horizon wavemode today. Splitting into short and long wavemode parts results in a splitting of the convolution integrals.
The non-Gaussian curvature perturbation is also split into $\zeta_{l}$ and $\zeta_{s}$. For scales well within the subvolume, the locally observed random field is $\zeta_{s}$; $\zeta_{l}$ is the scalar curvature perturbation smoothed over a scale $M$, so it is a background to fluctuations observed in a subsample of size $M$. As described in Section \ref{sec:LSsplit},
the local background $\zeta_l$ shifts the amplitude of the fluctuations as defined in the small volume, $\zeta\o = \zeta_s/(1+\zeta_l)$, but for our purposes it is safe to neglect this shift, so $\zeta\o\simeq\zeta_s$.
Carrying out the long- and short-wavelength split, we find\footnote{\label{footnoteIintegral}Note that splitting the momentum space convolution integrals yields factors of $\int \frac{d^3p}{(2\pi)^3} \sigma_{Gl}(\mathbf{p})$, which can be set equal to $\sigma_{Gl}(\mathbf{x}_0)=\int\frac{d^3p}{(2\pi)^3} \sigma_{Gl}e^{i\mathbf{p}\cdot\mathbf{x}_0}$ in the long-wavelength limit $\mathbf{p}\cdot\mathbf{x}_0\ll1$. Here, $\mathbf{x}_0$ is the location of the subvolume.}
\bea
\zeta\o_{\mathbf{k}} &=& \phi_{Gs,\mathbf{k}} + \left[1+\frac{6}{5}f_{\rm NL}(k)\sigma_{Gl}+\frac{27}{25}g_{\rm NL}(k)\sigma_{Gl}^2\right]\sigma_{Gs,\mathbf{k}} \nonumber \\
&&+\left[\frac{3}{5}f_{\rm NL}(k)+\frac{27}{25}g_{\rm NL}(k)\sigma_{Gl}\right](\sigma_{Gs}^2)_{\mathbf{k}}+\frac{9}{25}g_{\rm NL}(k) (\sigma_{Gs}^3)_{\mathbf{k}}+ \ldots,
\label{zetakobs}
\eea
where we have neglected corrections from quartic and higher terms. We see that the scale-dependent coefficients are corrected by long-wavelength pieces from higher terms and may scale differently in the small volume.

With the assumptions that the two fields are not correlated, $\Expect{\phi_{\mathbf{k}_1} \sigma_{\mathbf{k}_2}}=0$, and neglecting $g_{\rm NL}$ and higher order terms, the total power spectrum in the large volume  is
\bea\label{pzeta}
P_{\zeta}(k)&=&P_{\phi}(k)+P_{\sigma}(k)+\frac{18}{25}f_{\rm NL}^2\int_{L^{-1}}^{k_{\text{max}}}\frac{d^3p}{(2\pi)^3}P_{\sigma}(p)P_{\sigma}(|\mathbf{k}-\mathbf{p}|) \nn\\
&\simeq&P_{\phi}(k)+\Big(1+\frac{36}{25}f_{\rm NL}^2(k)\left(\langle\sigma_{Gl}^2\rangle+\langle\sigma_{Gs}^2(k)\rangle\right)\Big) P_{\sigma}(k),
\eea
where $P_{\phi}$ and $P_{\sigma}$ are the power spectra for the \textit{Gaussian} fields $\phi_G$ and $\sigma_G$, and we identify $\langle\sigma_{Gs}^2(k)\rangle = \int_{M^{-1}}^k \frac{d^3p}{(2\pi)^3}P_\sigma(p)$, $\langle\sigma_{Gl}^2\rangle = \int_{L^{-1}}^{M^{-1}} \frac{d^3p}{(2\pi)^3}P_\sigma(p)$, and for future use we define $\langle\sigma_{G}^2(k)\rangle = \langle\sigma_{Gl}^2\rangle+\langle\sigma_{Gs}^2(k)\rangle$. We also define $n_{\sigma}\equiv\frac{d\ln\mathcal{P}_{\sigma}}{d\ln k}$ and $n_{\phi}\equiv\frac{d\ln\mathcal{P}_{\phi}}{d\ln k}$ for future reference, where $\mathcal{P}_{\sigma}(k)\equiv\frac{k^3}{2\pi^2}P_{\sigma}(k)$ and $\mathcal{P}_{\phi}(k)\equiv\frac{k^3}{2\pi^2}P_{\phi}(k)$. In the second line of \eqref{pzeta} we have split the integral of the first line at the scale $M^{-1}$ after using the approximation \cite{Kumar:2009ge,Bramante:2011zr,Sugiyama:2012tr},
\be
\int_{L^{-1}}^{p_{\rm max}} \frac{ d^3 p}{ (2\pi)^3} P_{\sigma}(p)P_{\sigma}(|\mathbf{k}-\mathbf{p}|) \simeq 2 P_\sigma(k) \int_{L^{-1}}^k \frac{d^3p} {(2\pi)^3} P_\sigma(p). \label{leadinglog}
\ee
The fractional contribution of the non-Gaussian source to the total power is
\be
\label{eq:Definexi}
\xi_m(k)\equiv \frac{P_{\sigma, NG}(k)}{P_{\zeta}(k)}\;,
\ee
where, $P_{\sigma, NG} \equiv P_{\sigma}(k)+\frac{18}{25}f_{\rm NL}^2\int_{L^{-1}}^{k_{\text{max}}}\frac{d^3p}{(2\pi)^3}P_{\sigma}(p)P_{\sigma}(|\mathbf{k}-\mathbf{p}|)$ includes all contributions from the $\sigma$ sector of the perturbations. In the weakly non-Gaussian regime, $\xi_m(k)\approx P_{\sigma}(k)/P_{\zeta, G}(k)=P_{\sigma}(k)/(P_{\sigma}(k)+P_{\phi}(k))$. This ratio is a weakly scale-dependent function if the power spectra are not too different, so we parametrize it as
\be
\label{eq:xim}
\xi_m(k) = \xi_m(k_p)\left(\frac{k}{k_p}\right)^{n_f^{(m)}}\;.
\ee

Splitting off the long-wavelength background in Eq.~\eqref{zetakobs}, the curvature observed in a subvolume is
\bea\label{Pprime1}
P\o_{\zeta}(k)&=&P_{\phi}(k)+\Big(1+\frac{12}{5}f_{\rm NL}(k)\sigma_{Gl}+\frac{36}{25}f_{\rm NL}^2(k)\sigma_{Gl}^2\Big) P_{\sigma}(k) \nn\\
&&+ \frac{18}{25}f_{\rm NL}^2(k)\int_{M^{-1}}^{k_{\text{max}}}\frac{d^3p}{(2\pi)^3}P_{\sigma}(p)P_{\sigma}(|\mathbf{k}-\mathbf{p}|) \nn \\
&\simeq& P_{\phi}(k)+\Big(1+\frac{12}{5}f_{\rm NL}(k)\sigma_{Gl}+\frac{36}{25}f_{\rm NL}^2(k)\left(\sigma_{Gl}^2+\langle\sigma_{Gs}^2(k)\rangle\right)\Big) P_{\sigma}(k)
\eea
This expression shows that the local power on scale $k$ in typical subvolumes may be nearly Gaussian even if the global power on that scale is not. In other words, consider Eq.~(\ref{pzeta}) in the case that $\frac{36}{25}f_{\rm NL}^2(k)\langle\sigma_{Gl}^2\rangle >1$ and $\langle\sigma_{Gl}^2\rangle \gg \langle\sigma_{Gs}^2(k)\rangle $. The field $\sigma$ on scale $k$ is strongly non-Gaussian. However, in subvolumes with $\sigma_{Gl}^2\simeq\langle\sigma_{Gl}^2\rangle$ the contribution to the local power spectrum quadratic in $\sigma_{Gl}$ (the last term in the first line of Eq.~(\ref{Pprime1})) will give the dominant contribution to the Gaussian power while the local $f_{\rm NL}^2$ term (the term in the second line of Eq.~(\ref{Pprime1})) can be dropped. The locally observed $\sigma$ field on scale $k$ is weakly non-Gaussian.

When the locally observed field is nearly Gaussian (although the global field need not be, as described in the previous paragraph) the observed relative power of the two sources will vary in small volumes and is given by
\be
\xi\o_m(k)=\xi_m(k)\left[\frac{(1+\frac{6}{5}f_{\rm NL}(k)\sigma_{Gl})^2+\frac{36}{25}f_{\rm NL}^2(k)\langle\sigma^2_{Gs}(k)\rangle}{1+\frac{36}{25}f^2_{\rm NL}(k)\langle\sigma^2_{G}(k)\rangle+\frac{12}{5}\xi_m(k)f_{\rm NL}(k)\Big(\sigma_{Gl}+\frac{3}{5}f_{\rm NL}(k)(\sigma^2_{Gl}-\langle\sigma^2_{Gl}\rangle)\Big)}\right]\;.\label{xiobs}
\ee
Notice that  $\xi\o_m(k)|_{\sigma_{Gl}=0}=\xi_m(k)$, and that $\xi_m(k)=1$ implies $\xi_m\o(k)=1$.

\subsection{The bispectrum and the level of non-Gaussianity}
\label{subec:bispect}
From the generalized local ansatz in Eq.~(\ref{eq:GenLocalRealSpace}), the large volume bispectrum is
\be
\label{eq:BispectL}
B(k_1,k_2,k_3)=\frac{6}{5}f_{\rm NL}(k_3)\xi_m(k_1)\xi_m(k_2)P_{G,\zeta}(k_1)P_{G,\zeta}(k_2) + {\rm 2~perms.} + \mathcal{O}(f_{\,\rm NL}^3)+\dots
\ee
where the total Gaussian power, $P_{G,\zeta}$, comes from $\zeta_G\equiv\phi_G+\sigma_G$. The terms proportional to three or more powers of $f_{\rm NL}$ (evaluated at various scales) come both from the contribution from three copies of the quadratic $\sigma_G$ term from Eq.~(\ref{eq:GenLocalRealSpace}) and from the conversion between $P_{G,\sigma}$ and $P_{G,\zeta}$. Those terms may dominate the bispectrum if the model is sufficiently non-Gaussian over a wide enough range of scales. The same quantity as observed in a weakly non-Gaussian local subvolume is

\bea
\label{Bprime}
B\o_{\zeta}(k_1,k_2,k_3)&=&\frac{6}{5}\left[f_{\rm NL}(k_3)+\frac{9}{5}g_{\rm NL}(k_3)\sigma_{Gl}\right]\frac{\xi_m\o(k_1)P\o_{G,\zeta}(k_1)}{1+\frac{6}{5}f_{\rm NL}(k_1)\sigma_{Gl}+\frac{27}{25}g_{\rm NL}(k_1)\sigma_{Gl}^2}\\\nonumber
&&\times\frac{\xi_m\o(k_2)P\o_{G,\zeta}(k_2)}{1+\frac{6}{5}f_{\rm NL}(k_2)\sigma_{Gl}+\frac{27}{25}g_{\rm NL}(k_2)\sigma_{Gl}^2}+ {\rm 2~perms.}+\dots\\\nonumber
\eea
where the $\dots$ again denote terms proportional to more power of $f_{\rm NL}$. Comparing this expression to the previous equation in the weakly non-Gaussian regime, the observed bispectrum is again a product of functions of $k_1$, $k_2$, and $k_3$. However, those functions are no longer equivalent to the Gaussian power and ratio of power in the two fields that would be measured from the two-point correlation. In other words, the coupling to the background not only shifts the amplitude of non-Gaussianity, but can also introduce new $k$-dependence which alters the shape of the small-volume bispectrum. Although a full analysis of a generic local type non-Gaussianity would be very useful, for the rest of this section we set $g_{NL}$ and all higher terms to zero for simplicity.

These bispectra now have a more complicated shape than in the standard local ansatz, but for weak scale-dependence they are not still not too different. In practice one defines an $f_{\rm NL}$-like quantity from the squeezed limit of the bispectrum:
\be\label{fNLS}
\frac{3}{5}f^{\rm eff}_{\rm NL}(k_s, k_l)\equiv\frac{1}{4}\lim_{k_l\rightarrow0}\frac{B\o_{\zeta}(\mathbf{k}_l,\mathbf{k}_s,-\mathbf{k}_l-\mathbf{k}_s)}{P\o_{\zeta}(k_l)P\o_{\zeta}(k_s)},
\ee
where $P\o_{\zeta}$ and $B\o_{\zeta}$ are defined in terms of $\zeta\o$.  The definition of $f^{\rm eff}_{\rm NL}$ in Eq.~\eqref{fNLS} is imperfect in any finite volume, since we cannot take the exact limit $k_l\rightarrow0$. Instead, we must choose the long and short wavelength modes ($k_l\ll k_s$) from within some range of observable scales.

Since the best observational constraints over the widest range of scales currently come from the CMB, we will fix $k_l$ and $k_s$ in terms of the range of angular scales probed by \textit{Planck} and define
\be
\label{eq:fnlCMB}
 f^{\rm CMB}_{\rm NL}(k_s,k_l)\equiv f^{\rm eff}_{\rm NL}(k_s,k_l)|_{k_s=k_{\rm CMB\, max},\,k_l=k_{\rm CMB\, min}}\;.
\ee
The observed non-Gaussianity in a subvolume can then be expressed in terms of the large volume quantities as
\be\label{fNLS2}
 f^{\rm CMB}_{\rm NL}=\frac{f_{\rm NL}(k_s)\xi_m(k_s)\xi_m(k_l)\Big(1+\frac{6}{5}f_{\rm NL}(k_s)\sigma_{Gl}\Big)\Big(k_s\rightarrow k_l\Big)}{\Big[1+\frac{36}{25}f^2_{\rm NL}(k_s)\langle\sigma^2_{G}(k_s)\rangle+\frac{12}{5}\xi_m(k_s)f_{\rm NL}(k_s)\Big(\sigma_{Gl}+\frac{3}{5}f_{\rm NL}(k_s)(\sigma^2_{Gl}-\langle\sigma^2_{Gl}\rangle)\Big)\Big]\Big[k_s\rightarrow k_l\Big]},
\ee
where $(k_s \rightarrow k_l)$ indicates the same term as the preceding, except with $k_s$ replaced by $k_l$, and this expression is evaluated with $k_l$, $k_s$ equal to the limiting wavemodes observed in the CMB. As in the discussion below Eq.~(\ref{Pprime1}), this expression is valid even for $f_{\rm NL}(k)\sigma_{Gl}\gtrsim1$ as long as we can neglect the $1$-loop contribution to $P\o_{\zeta}$; this must always be the case for our observed universe with very nearly Gaussian statistics. Keep in mind that $ f^{\rm CMB}_{\rm NL}$ is \textit{not} a small volume version of the parameter $f_{\rm NL}(k)$ defined in Eqs. \eqref{eq:GenLocalRealSpace}, \eqref{zetak}, which is a function of a single scale.  Rather, $ f^{\rm CMB}_{\rm NL}$ corresponds to the observed amplitude of nearly local type non-Gaussianity over CMB scales for given values of $f_{\rm NL},\xi_m,k_s,k_l,\sigma_{Gl}$.

In the discussions above, we have considered the case when modes of a particular scale $k$ may be strongly coupled (the term quadratic in $f_{\rm NL}(k)$ dominates in $P(k)$). However, it is also useful to have a measure of total non-Gaussianity that integrates the non-Gaussian contributions on all scales. For this we use the dimensionless skewness
\be
\mathcal{M}_3\equiv\frac{\langle\zeta^3(\mathbf{x})\rangle}{\langle\zeta^2(\mathbf{x})\rangle^{3/2}}\ll1. \label{M3def}
\ee
There are two important things to notice about this quantity compared to $f_{\rm NL}(k)$ in the local ansatz itself and $f^{\rm CMB}_{\rm NL}(k_s,k_l$) as defined in Eq.~(\ref{fNLS2}).

First, $f_{\rm NL}(k)\langle\sigma_{Gl}^2\rangle^{1/2}\gtrsim1$ does not necessarily imply $\mathcal{M}_3>1$, even in the single source case. The behavior of the power spectrum and bispectrum over the entire span of superhorizon and subhorizon e-folds enter $\mathcal{M}_3$.
Evaluating $\mathcal{M}_3\equiv\frac{\langle\zeta^3(\mathbf{x})\rangle}{\langle\zeta^2(\mathbf{x})\rangle^{3/2}}$ in the single-source case ($\sigma_G\rightarrow\zeta_G$, $n_{\sigma}\rightarrow n_{\zeta}$, as defined in Section~\ref{sec:LSsplit}) for a scale-dependent scalar power spectrum, $n_{\zeta} \neq 1$, yields
\bea
\label{M3allns}
\mathcal{M}_3 &\simeq& \frac{36}{5}f_{\rm NL}(k_l)\langle\zeta_{Gl}^2\rangle^{1/2}\left(1+\frac{e^{N_{\rm sub}(n_s-1)}-1}{1-e^{-N(n_{\zeta}-1)}} \right)^{1/2}\frac{(n_{\zeta}-1)e^{-N(n_f+2n_{\zeta}-2)}}{\left(1-e^{-(N+N_{\rm sub})(n_{\zeta}-1)}\right)^2}\nonumber \\ && \times \left[\frac{e^{(N+N_{\rm sub})(n_f+2n_{\zeta}-2)}-1}{(n_f+2n_{\zeta}-2)}-\frac{e^{(N+N_{\rm sub})(n_f+n_{\zeta}-1)}-1}{(n_f+n_{\zeta}-1)}\right],
\eea
where we have used the approximation shown in Eq.~(\eqref{leadinglog}), and neglected 1-loop and higher contributions to $\langle\zeta^3\rangle$ and $\langle\zeta^2\rangle$. Splitting the total e-folds on a scale appropriate for our cosmology, the number of subhorizon e-folds is $N_{\text{sub}}=60$. For a scale-invariant spectrum, $n_{\zeta}=1$, the expression above becomes
\bea
\label{M3ns1}
\mathcal{M}_3&\simeq&\frac{36}{5}f_{\rm NL}(k_l)\langle\zeta_{Gl}^2\rangle^{1/2}\left(1+ \frac{N_{\rm sub}}{N} \right)^{1/2}\frac{1}{n_f^2(N+N_{\rm sub})^2}\\\nonumber
&&\times
\left[e^{n_fN_{\text{sub}}}(-1+n_f(N+N_{\text{sub}}))+e^{-n_fN}\right].
\eea
For the scale-independent case, $n_f=0$, this reduces to
\be
\mathcal{M}_3=\frac{18}{5}f_{\rm NL}\mathcal{P}_{\zeta}^{1/2}(N+N_{\rm sub})^{1/2}\;.\label{M3scaleinv}
\ee
Notice that for $n_{\zeta}=1$ and $n_f<0$ (the case of increasing non-Gaussianity in the IR), $\mathcal{M}_3$ grows rapidly with $N$. On the other hand, if the power spectrum has a red tilt, $n_{\zeta} < 1$, $\mathcal{M}_3$ will stay small for a wider range of $n_f$ values.

The second thing to keep in mind about the $\mathcal{M}_n$ is that the series gives a more accurate characterization of the total level of non-Gaussianity than $f^{\rm CMB}_{\rm NL}$ or $\mathcal{M}_3$ alone would. The level of non-Gaussianity as determined by $\mathcal{M}_{n+1}/\mathcal{M}_n$ is also what controls the size of the shift small volume quantities can have due to mode coupling. For example, in the two-field case the quantity controlling the level of non-Gaussianity of $\zeta$ is $\xi_mf_{\rm NL}\mathcal{P}_{\zeta}^{1/2}$, where $\xi_m(k)$ is the fraction of power coming from
$\sigma_G$ in the weakly non-Gaussian case. This quantity determines the scaling of the dimensionless non-Gaussian cumulants,
\be
\mathcal{M}_n\equiv\frac{\langle\zeta(\mathbf{x})^n\rangle_c}{\langle\zeta(\mathbf{x})^2\rangle^{n/2}}
\propto\left.\xi_m\left[\xi_mf_{\rm NL}\mathcal{P}_{\zeta}^{1/2}\right]^{n-2}\right|_{k_p}.
\ee
(We specify the scale-dependent functions at some pivot scale as the cumulants involve integrals over these functions at all scales.) The quantity $\xi_mf_{\rm NL}\mathcal{P}_{\zeta}^{1/2}$ as defined in a subvolume differs from the large-volume quantity due to coupling to background modes \cite{LoVerde:2013xka}:
\be
\left.\xi_mf_{\rm NL}\mathcal{P}_{\zeta}^{1/2}\right|_{\text{obs}}=\xi_mf_{\rm NL}\mathcal{P}_{\zeta}^{1/2}\left[1-\frac{6}{5}\xi_mf_{\rm NL}\langle\zeta_G^2\rangle B\right],
\ee
where we have suppressed the scale-dependence, and the bias is now defined as
\be
B\equiv\sigma_{Gl}/\langle\zeta_G^2\rangle^{1/2}, \label{biastwofield}
\ee
so that it is larger when $\sigma$, which biases the subsamples, is a larger fraction of the curvature perturbation.

\section{Observational consequences}
\label{sec:ObsConsequences}
In this section we illustrate the range of large-volume statistics that can give rise to locally observed fluctuations consistent with our observations. In considering the relationship between {\it Planck} CMB data and inflation theory, we set the scale of the subvolume to be $M\approx H_0^{-1}$.

\subsection{The shift to the power spectrum}
\label{subsec:power}

Expressed in terms of the large volume power spectrum Eq.~(\eqref{pzeta}), the small volume power spectrum Eq.~\eqref{Pprime1} is
\be
P\o_{\zeta}(k)=P_{\zeta}(k)\left[1+\frac{12}{5}\xi_m(k)\frac{f_{\rm NL}(k)\sigma_{Gl}+\frac{3}{5}f_{\rm NL}^2(k)(\sigma^2_{Gl}-\langle\sigma^2_{Gl}\rangle)}{1+\frac{36}{25}f_{\rm NL}^2(k)\langle\sigma_{G}^2(k)\rangle}\right], \ \
\label{Pprime}
\ee
In the single field, scale-independent, weakly non-Gaussian limit, $\xi_m=1$ and $f_{\rm NL}=const.$, and Eq.~\eqref{Pprime} reduces to Eq.~\eqref{Pobs}.
The shift to the local power spectrum is proportional to the level of non-Gaussianity $\xi_m(k) f_{\rm NL}(k)\langle\zeta_G^2\rangle^{1/2}$ coupling subhorizon modes to long-wavelength modes. We will see in Section~\ref{sec:spectralindex} below that if mode coupling is weaker on superhorizon scales, $\xi_m(k)f_{\rm NL}(k)\langle\zeta_G^2\rangle^{1/2}\gtrsim1$ can be consistent with weak global non-Gaussianity.

Depending on the value of $\xi_m(k)$ and on the biasing quantity $\frac{6}{5} f_{\rm NL}(k)\sigma_{Gl}$ on the scale $k$, this shift is approximately
\be
\label{Pobslimits}
\frac{P\o_{\zeta}}{P_{\zeta}}  \approx \left\{
  \begin{array}{ll}
1+\frac{12}{5} \xi_m f_{\rm NL}\sigma_{Gl}, & \hspace{5mm} \frac{6}{5}f_{\rm NL}(k)\sigma_{Gl}\ll1 \\
1+\xi_m\Big(\frac{\sigma_{Gl}^2-\langle\sigma_{Gl}^2\rangle}{\langle\sigma_{G}^2(k)\rangle}\Big), & \hspace{5mm} \frac{6}{5}f_{\rm NL}(k)\sigma_{Gl}\gg1
  \end{array}
 \right.
\ee
In the $\frac{6}{5}f_{\rm NL}(k)\sigma_{Gl}\ll1$ limit, the shift to the observed power comes from the $\mathcal{O}(f_{\rm NL}\sigma_{Gl})$ term, which increases or decreases the power from the field $\sigma$. In addition, the spectral index can change if the non-Gaussianity is scale-dependent (note the additional $k$-dependence from the $f_{\rm NL}(k)\sigma_{Gl}$ term in Eq.~\eqref{Pprime1} as compared to Eq.~\eqref{pzeta}). New scale dependence can also be introduced if there are two sources contributing to $\zeta$ and one is non-Gaussian. In the $\frac{6}{5}f_{\rm NL}(k)\sigma_{Gl}\gg 1$ limit, where the global power $P_{\sigma}(k)$ on subhorizon scales is dominated by the 1-loop contribution, the $\mathcal{O}(f_{\rm NL}^2\sigma_{Gl}^2)$ term dominates. If the size of the background fluctuation is larger (smaller) than $1\sigma$, the power from the field $\sigma$ will be increased (decreased) relative to the global average,\footnote{Note that there is a decrease in power in the majority of subvolumes, which is balanced by a strong increase in power in more rare subvolumes, so that the average power over all subvolumes recovers the large-volume power, $\langle P\o_{\zeta}\rangle_{M\in L}=P_{\zeta}$.} but with the same scale-dependence.
Consequently, a shift in $n_s$ comes from the difference in running between the two fields: the observed running $n\o_s$ will be shifted by the running of the fields $\phi_G$ or $\sigma$, depending on whether the power from the field $\sigma$ is increased or decreased (see Eq.~\eqref{eq:Deltans} below). Alternatively, if $f_{\rm NL}(k)\sigma_{Gl}=\mathcal{O}(1)$ on or near observable scales, $n_s$ can be shifted due to the relative change in power of the linear and quadratic pieces of $\sigma$; this scenario is shown below in Figure~\ref{SingleSourcePowerShift}.

\subsection{The shift to the spectral index, $\Delta n_s$}
\label{sec:spectralindex}

Eq.~\eqref{Pprime} shows that the presence of a superhorizon mode background causes the spectral index $\frac{d\ln\mathcal{P}_{\zeta}}{d\ln k}\equiv n_s-1$ to vary between subvolumes.\footnote{Recall that $n_s$ is the running of the total field, in contrast with $n_{\sigma,\phi}$ ($n_{\zeta}$) for the Gaussian fields $\sigma_G$, $\phi_G$ ($\zeta_G$) in the multi-source (single-source) case.} Taking the logarithmic derivative of Eq.~(\ref{Pprime}) with respect to $k$, we find 
\bea
 \Delta n_{s}(k)& \equiv& n\o_s-n_s \nonumber
\\ &=&\frac{\frac{12}{5} \xi_m f_{\rm NL}\left(\sigma_{Gl}
(n_f^{(m)}+X_1n_f)+\frac{3}{5}f_{\rm NL}(\sigma_{Gl}^2-\langle \sigma_{Gl}^2\rangle)(n_f^{(m)}+X_2n_f)\right)}{1+\frac{36}{25}f_{\rm NL}^2\langle\sigma_{G}^2(k)\rangle+\frac{12}{5}\xi_mf_{\rm NL}\Big(\sigma_{Gl}+\frac{3}{5}f_{\rm NL}(\sigma_{Gl}^2-\langle \sigma_{Gl}^2\rangle)\Big)}, \ \ \ \ \ \
\label{nsprime}
\eea
where from Eq.~(\ref{eq:nfng}) and Eq.~(\ref{eq:xim}), $n_f\equiv\frac{d\ln f_{\rm NL}}{d\ln k}$, $n_f^{(m)}\equiv\frac{d\ln\xi_m}{d\ln k}$,
\be
X_1\equiv\frac{1-\frac{36}{25}f_{\rm NL}^2\langle\sigma_{G}^2(k)\rangle}{1+\frac{36}{25}f_{\rm NL}^2\langle\sigma_{G}^2(k)\rangle}, \ \ \ \ \ X_2\equiv\frac{2}{1+\frac{36}{25}f_{\rm NL}^2\langle\sigma_{G}^2(k)\rangle},\nn
\ee
and we have mostly suppressed the $k$-dependence.
From either \eqref{Pobslimits} or \eqref{nsprime} we see that depending on the value of $\xi_m(k)$ and on the level of non-Gaussianity $\frac{6}{5} f_{\rm NL}(k)\sigma_{Gl}$ on the scale $k$, this shift is approximately
\bea
\Delta n_s&\approx&
\left\{
  \begin{array}{ll}
\frac{12}{5} \xi_m f_{\rm NL}\sigma_{Gl}
(n_f^{(m)}+n_f), & \hspace{5mm} \frac{6}{5}f_{\rm NL}(k)\sigma_{Gl}\ll1 \\
n_f^{(m)}\left(\frac{\sigma_{Gl}^2-\langle\sigma_{Gl}^2\rangle}{\xi_m^{-1}\langle\sigma_{G}^2(k)\rangle + \sigma_{Gl}^2-\langle\sigma_{Gl}^2\rangle}\right), & \hspace{5mm} \frac{6}{5}f_{\rm NL}(k)\sigma_{Gl}\gg 1
  \end{array}
 \right.
 \label{eq:Deltans}
\eea
where these expressions are approximate, and in particular the single-source limit $\frac{6}{5}f_{\rm NL}(k)\sigma_{Gl}\gg 1$ cannot be taken simply as the $n_f^{(m)}\rightarrow0, \ \xi_m\rightarrow1$ limit of Eq.~\eqref{eq:Deltans}. That limit requires the full expression, \eqref{nsprime}, from which we find that in the single source case when $\frac{6}{5} f_{\rm NL}(k)\sigma_{Gl}\gg1$ and $\sigma_{Gl}^2\gg\langle\sigma_{Gs}^2(k)\rangle$, the correction to the power spectrum vanishes, $\Delta n_s \simeq - n_f / (\frac{3}{5} f_{\rm NL}(k)\sigma_{Gl}) \rightarrow 0$. This equation indicates that these scenarios will also in general have a non-constant spectral index. Although we have not done a complete analysis, Eq.(\ref{nsprime}) shows that $\alpha\o_s(k) \equiv d\,{\rm ln}\, n\o_s/d\, {\rm ln}\, k \neq \alpha_s(k)$ should generically be of order slow-roll parameters squared, which is consistent with {\it Planck} results \cite{Ade:2013uln}. 

The shift to the spectral index is thus determined by runnings in the large-volume bispectrum, the level of non-Gaussianity on scale $k$ (the strength of mode coupling between this scale and larger scales), and the amount of bias for the subvolume, which will depend on the number of superhorizon e-folds along with the size and running of the power spectrum outside the horizon. We stress that this shift depends on the non-Gaussianity and non-Gaussian running \emph{of the statistics at the scale being measured}, and does not depend directly on the superhorizon behavior of the bispectrum parameters $f_{\rm NL}(k)$, $\xi_m(k)$. We will see below that even if $f_{\rm NL}(k)$ or $\xi_m(k)$ fall swiftly to zero outside the observable volume, the shift $\Delta n_s$ will be significant if subhorizon modes $k>H_0^{-1}$ are strongly coupled to superhorizon modes.

Note also that the bias from a given background mode does not depend on the scale of the mode (except through the scale-dependence of $\mathcal{P}_{\sigma}$) as $\sigma_{Gl}$ simply adds up all the background modes equally. We will see in Section \ref{sec:NonLocalBispect} that this is not true for nonlocal mode coupling: infrared modes of different wavelength can be weighted differently.

For the purpose of model building, it should be pointed out that when $\frac{6}{5} f_{\rm NL} \sigma_{Gl} <0$, equations \eqref{xiobs} and \eqref{nsprime} can diverge. For instance for a single source model with a hundred superhorizon e-folds ($\langle\zeta_{Gs}^2\rangle\sim0$), equation \eqref{nsprime} is inversely proportional to factors of $(1+ \frac{6}{5} f_{\rm NL} \sigma_{Gl})^2$. This would be cause for concern -- naively it implies extremely large corrections to the spectral index when $\frac{6}{5} f_{\rm NL} \sigma_{Gl}\sim -1$. However because of the same proportionality, Eq. \eqref{fNLS2} will also diverge, indicating that the subsamples in this phase space would observe extremely non-Gaussian statistics ($f_{\rm NL}\o \gg 10$). Hence the $\textit{Planck}$ satellite's bound on non-Gaussianity has already excluded the worst-behaved phase space for a negative combination of parameters and background fluctuation, $\frac{6}{5} f_{\rm NL} \sigma_{Gl} = \frac{6}{5} f_{\rm NL} \langle \sigma_{Gl}^2 \rangle^{1/2}B < 0$.

It was shown in \cite{Nelson:2012sb,Nurmi:2013xv,LoVerde:2013xka} that strong non-Gaussianity in a large volume can be consistent with weak non-Gaussianity measured in typical subvolumes. Furthermore, for scale-dependent non-Gaussianity, large $f_{\rm NL}(k)$ on a given scale can be consistent with weak total non-Gaussianity (adding over all scales). In light of this, we would like to better understand for what values of the parameters, and in particular the global spectral index and bispectral indices, it is possible for a shift $|\Delta n_s|\sim 0.04$ to be typical in Hubble-sized subvolumes, while satisfying the following theoretical and observational conditions:

\begin{enumerate}
\item $\zeta$ is a small perturbation. We will impose this by requiring that the amplitude of fluctuations is small for each term in the local ansatz. \label{zetasmall}
\item The observed power spectrum $\mathcal{P}\o_{\zeta}(k_p)=2.2\times10^{-9}$, where $k_p=0.05\text{ Mpc}^{-1}$ \cite{Ade:2013ktc}, should be typical for subvolumes. We will enforce this condition by setting $\mathcal{P}\o_{\zeta}(k_p)$ as given in Eq.~\eqref{Pprime} equal to the power in a subvolume with a typical background fluctuation.
This determines the number of superhorizon e-folds, $N$, in terms of $n_{\zeta}$, $f_{\rm NL}(k_p)$, and $\langle\sigma_{Gl}^2\rangle$ in the case of single-source perturbations, while for two sources a choice of $\xi_m(k_p)$ is also needed to fix $N$.

\item The observed level of non-Gaussianity in typical Hubble-sized subvolumes is consistent with {\it Planck} satellite bounds.
Using Eq.~(\ref{eq:fnlCMB}) we require $f^{\rm CMB}_{\rm NL}\leq10$ for a typical background fluctuation, although a more precise analysis could be done. The maximum and minimum multipoles $(l_{\text{max}},l_{\text{min}})=(2500,1)$ used to estimate $f^{\rm CMB}_{\rm NL}$ in \cite{Ade:2013tta} translate into 3-dimensional wavenumbers $k_{\text{max}}=0.2\text{ Mpc}^{-1}$, $k_{\text{min}}=10^{-4}\text{ Mpc}^{-1}$ \cite{Ade:2013uln}, in Eq. \eqref{fNLS2}.

\item The total non-Gaussianity is weak, $\mathcal{M}_3\ll 1$. Our formulae are strictly correct for scenarios where the large volume is weakly non-Gaussian on all scales, {\it and} when some scales in the large volume are strongly coupled, but in typical subvolumes weakly non-Gaussian statistics are observed. To give some sense of the regime in which our expressions are not exact, our plots will indicate the parameter ranges where the total non-Gaussianity summed over all scales is not small, $\mathcal{M}_3\geq1$. If smaller modes are more strongly coupled, $n_f>0$, this constraint is generally weaker than the requirement of matching the observational constraints on non-Gaussianity. However, if the long wavelength modes are strongly coupled, $n_f<0$, this restriction can be quite important. 
\end{enumerate}

A further possible criteria might be to require $|\Delta n_s|\lesssim 0.1$; for larger values the observed near scale-invariance $n\o_s\simeq1$ might be an unlikely accident given the large variation in scale-dependence among subvolumes. However, Eq.~\eqref{eq:Deltans} shows that this condition is satisfied even for large $f_{\rm NL}(k)\langle\sigma_{Gl}^2\rangle^{1/2}$ as long as the non-Gaussian runnings $n_f$, $n_f^{(m)}$ are not too large, which is also necessary to preserve conditions 1 and 4 above. \\

\hspace{-25pt}\textbf{Example I: Single source perturbations with constant $f_{NL}$.}

\begin{figure}
\caption{Parameter space for single-source, scale-invariant non-Gaussianity. The shaded region on the right is marked off by lines where $\langle\zeta_{Gl}^2\rangle=0.1$ and $\langle[\frac{3}{5}f_{\rm NL}(\zeta_{Gl}^2-\langle\zeta_{Gl}^2\rangle)]^2\rangle=0.1$. The shaded region on the left shows the constraint on non-Gaussianity from \textit{Planck}; outside this region, $f^{\rm CMB}_{\rm NL}=f_{\rm NL}/(1+\frac{6}{5}f_{\rm NL}\zeta_{Gl})^2<10$ in subvolumes with a $+1\sigma$ background fluctuation $\zeta_{Gl}=\langle\zeta_{Gl}^2\rangle^{1/2}$ (for a $-1\sigma$ fluctuation, the constraint is similar but stronger). The dashed black line denotes $\mathcal{M}_3\simeq\frac{18}{5}f_{\rm NL}\langle\zeta_{G}^2\rangle^{1/2}=1$, dividing the weakly and strongly non-Gaussian regions (here we take $n_{\zeta}=1$). The dotted lines, from left to right, denote curves of constant $N=350$ for $n_{\zeta}=1.04$, $1$, and $0.96$.}
\centering
\includegraphics[width=.7\textwidth]{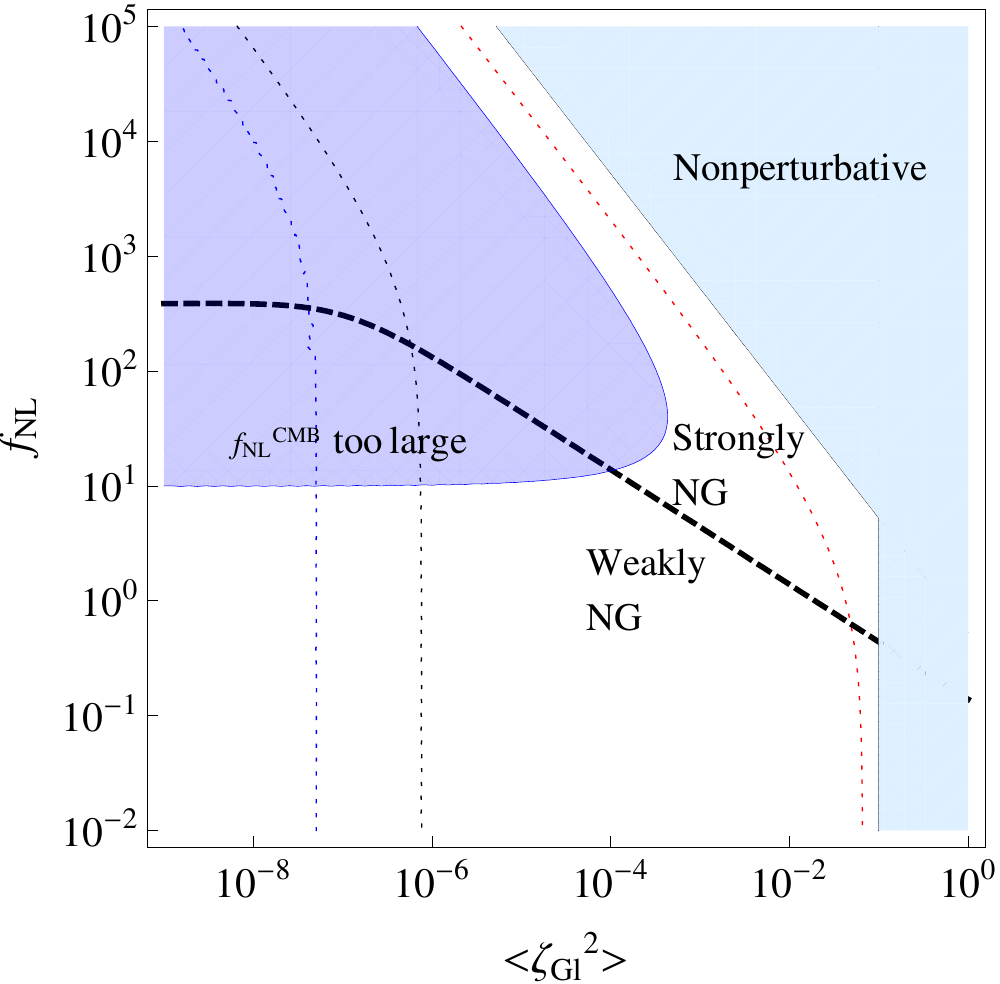}
\label{ConstraintsScaleInv}
\end{figure}

To understand how the conditions above affect the parameter space, consider first the simple case of single-source, scale-invariant non-Gaussianity with only $f_{\rm NL}$ non-zero:
\be
\label{eq:zetaagain}
\zeta=\zeta_G+\frac{3}{5}f_{\rm NL}(\zeta_G^2-\langle\zeta_G^2\rangle),
\ee
where $f_{\rm NL}$ is constant. In Figure~\ref{ConstraintsScaleInv}, we show the parameter space $(\langle\zeta_{Gl}^2\rangle,f_{\rm NL})$ consistent with $\zeta\ll1$, the observed power spectrum and observed bounds on non-Gaussianity. The dashed black line divides the parameter space where the entire volume is on average weakly or strongly non-Gaussian by setting $\mathcal{M}_3\simeq\frac{18}{5}f_{\rm NL}\langle\zeta_{G}^2\rangle^{1/2}=1$. This dashed line levels off in parameter space with very small superhorizon contributions to $\mathcal{M}_3$, $\langle\zeta_{Gl}^2\rangle\ll\langle\zeta_{Gs}^2\rangle$ (meaning subhorizon fluctuations dominate the cumulative skewness), which for a nearly scale-invariant power spectrum implies $N \ll N_{\rm sub}$. (Here and in the rest of this Section, we set the number of subhorizon e-folds $N_{\text{sub}}=60$.) When $\mathcal{M}_3\gtrsim\mathcal{O}(1)$, the dominant contribution to bispectrum in the large volume is given by the higher order terms not explicitly written in Eq.~(\ref{eq:BispectL}).

The shaded region to the right of the thin gray solid lines shows where $\zeta$ is no longer a small perturbation, either due to a large linear or quadratic term. The shaded region on the left shows where $f^{\rm CMB}_{\rm NL}$ in typical subvolumes is inconsistent with constraints from \textit{Planck}. We see that in the weakly non-Gaussian regime, consistency with \textit{Planck} reduces to $f_{\rm NL}< 10$, whereas in the strongly non-Gaussian regime the amplitude of fluctuations must be large enough, $\langle\zeta_l^2\rangle^{1/2}\sim f_{\rm NL}\langle\zeta_{Gl}^2\rangle\gtrsim\frac{1}{10}$, to sufficiently bias Hubble-sized subvolumes so that weak non-Gaussianity is typical. In this regime there is only a small window where 1$\sigma$ fluctuations give subvolumes consistent with observation, and requiring $f^{\rm CMB}_{\rm NL}$ to be a factor of 10 smaller would essentially remove this small window. This is because $f^{\rm CMB}_{\rm NL}\sim1/f_{\rm NL}\zeta_{Gl}^2\sim1/\zeta_l$, so if $f^{\rm CMB}_{\rm NL}$ is constrained to $\mathcal{O}(1)$, $\zeta$ is forced to be nonperturbative. Thus, for strongly non-Gaussian, scale-invariant superhorizon perturbations on a homogeneous background geometry to be consistent with observation, the degree of non-Gaussianity in our subvolume would have to exceed the observed degree of inhomogeneity, 1 part in $10^5$.

The remaining lines denote curves of constant $N=350$ for different values of $n_{\zeta}$ (which we take to be constant), fixed by the requirement that the observed amplitude of fluctuations be typical of subvolumes: $\mathcal{P}\o_{\zeta}(k_p)=(1+\frac{6}{5}f_{\rm NL}\zeta_{Gl})^2\langle\zeta_{Gl}^2\rangle\frac{n_{\zeta}-1}{1-e^{-N(n_{\zeta}-1)}}=2.2\times10^{-9}$ for a typical $+1\sigma$ background fluctuation ($\zeta_{Gl} =
\langle \zeta_{Gl}^2 \rangle^{1/2}$). The entire unshaded parameter space is consistent with the observed amplitude of fluctuations, once we impose this relationship between $N$ and the parameters plotted. For $f_{\rm NL}\langle\zeta_{Gl}^2\rangle^{1/2}\ll1$, $\mathcal{P}\o_{\zeta}\approx\mathcal{P}_G=\langle\zeta_{Gl}^2\rangle\left(\frac{n_{\zeta}-1}{1-e^{-N(n_{\zeta}-1)}}\right)$ is fixed by the observed power spectrum, so curves of constant $N$ approach a fixed value of $\langle\zeta_{Gl}^2\rangle$. On the other hand, for $f_{\rm NL}\langle\zeta_{Gl}^2\rangle^{1/2}\gg1$, $\mathcal{P}\o_{\zeta}\propto f_{\rm NL}^2\zeta_{Gl}^2\mathcal{P}_G$ and curves of constant $N$ approach lines of constant $f_{\rm NL}\langle\zeta_{Gl}^2\rangle$. The variation with $n_{\zeta}$ shows that for a red (blue) tilt, a given number of superhorizon e-folds corresponds to a much larger (smaller) amplitude of fluctuations $\langle\zeta_{Gl}^2\rangle$. For a red tilt or flat spectrum, there is a maximum number of e-folds consistent with $\langle\zeta_{Gl}^2\rangle<1$, whereas for a blue tilt as small as $n_\zeta-1\sim\mathcal{P}\o_\zeta\sim10^{-9}$,  $\langle\zeta_{Gl}^2\rangle$ will remain perturbatively small for an arbitrarily large number of e-folds. For instance, for $n_{\zeta}=1.04$, having more than 50 superhorizon e-folds does not appreciably change the value of $\langle\zeta_{Gl}^2\rangle$ in the region where $f_{\rm NL}\langle\zeta_{Gl}^2\rangle^{1/2}\ll1$ (the vertical part of the blue dashed line in Figure \ref{ConstraintsScaleInv} will not shift right with the addition of more superhorizon e-folds).

Note that, in the case where $f_{\rm NL}$ is scale-invariant, $\mathcal{M}_3$ is a function of superhorizon e-folds $N$ (Eq. \eqref{M3scaleinv}). In order to calculate the dashed line for fixed $\mathcal{M}_3$ in Figure \ref{ConstraintsScaleInv} we assume $n_{\zeta}=1$, which along with Eq.~\eqref{zetaGl^2} fixes the number of superhorizon e-folds in terms of $f_{\rm NL}$ and $\langle \zeta_{Gl}^2 \rangle$. In the strongly non-Gaussian regime, moving along the allowed window in parameter space (along curves of constant $N$) does not change the amplitude of fluctuations $\langle\zeta^2\rangle$ or statistics $\zeta\propto\zeta_G^2$, but only gives a relative rescaling to $f_{\rm NL}$ and $\langle\zeta_{Gl}^2\rangle$. That is, requiring weak non-Gaussianity of a given size in typical subvolumes from a strongly non-Gaussian large volume singles out (in the scale-invariant case) a particular amplitude of fluctuations in the large volume, and as described above, this amplitude becomes nonperturbative when $f\o_{\rm NL}\sim1$ is typical. In the following section we will see how this condition can be removed in the case of scale-dependent non-Gaussianity: a blue running of $f_{\rm NL}$ implies the level of non-Gaussianity attenuates at large scales.\\ 

\hspace{-25pt}\textbf{Example II: Single source with running $f_{\rm NL}(k)$.}

\begin{figure}
\caption{Parameter space for single source non-Gaussian models with $n_f=0.1$ in the upper panels and $n_f=-0.1$ in the lower panels. Left and right panels show parameter space for globally flat and red spectral indices, $n_{\zeta} = 1,~0.96$. The solid black lines show $\Delta n_s=- 0.04$ for $+0.5\sigma$ background fluctuations and positive $f_{\rm NL}$ (or $-0.5\sigma$ background fluctuations and negative $f_{\rm NL}$). The dotted-dashed lines indicate where $f_{\rm NL}(k_p) \langle \zeta_{Gl}^2 \rangle^{1/2} = 10$, above which $\Delta {n_s}$ will approach zero and $n_s(k_p) \simeq n_\zeta + 2 n_f$. The far right region, $\langle \zeta_{Gl}^2 \rangle\gtrsim 0.1$, is nonperturbative, along with the nonperturbative region $\langle[\frac{3}{5}f_{\rm NL} \star(\zeta_{Gl}^2-\langle\zeta_{Gl}^2\rangle)]^2 \rangle\gtrsim 0.1$, which excludes parameter space for a red tilt of $f_{\rm NL}(k)$ ($n_f<0$). The upper left regions show the observational constraint $f^{\rm CMB}_{\rm NL}<10$ from Planck. The dashed curves show $\mathcal{M}_3\simeq1$, and thus divide weakly and strongly non-Gaussian parametrizations. The dotted lines indicate how many superhorizon e-folds are implied by the choice of $n_f, n_{\zeta},\langle \zeta_{Gl}^2 \rangle,$ and $f_{\rm NL}(k_p)$. As discussed after Eq.~\eqref{nsprime} and indicated in the upper right of the top panels, $\frac{6}{5}f_{\rm NL}(k_p)\langle \zeta_{Gl}^2 \rangle^{1/2} \gg 1$ implies $\Delta n_s \rightarrow 0$. The black squares mark phase space for $\Delta n_s$ probabilities plotted in Figure \ref{fig:ProbDeltans}.}
\centering
\begin{tabular}{cccc}
$\bf{n_{\zeta}=1,~n_f=0.1}$ &\hspace{.7 cm} $\bf{n_{\zeta}=0.96,~n_f=0.1}$ \\
\hspace{-.5cm}\includegraphics[width=.49\textwidth]{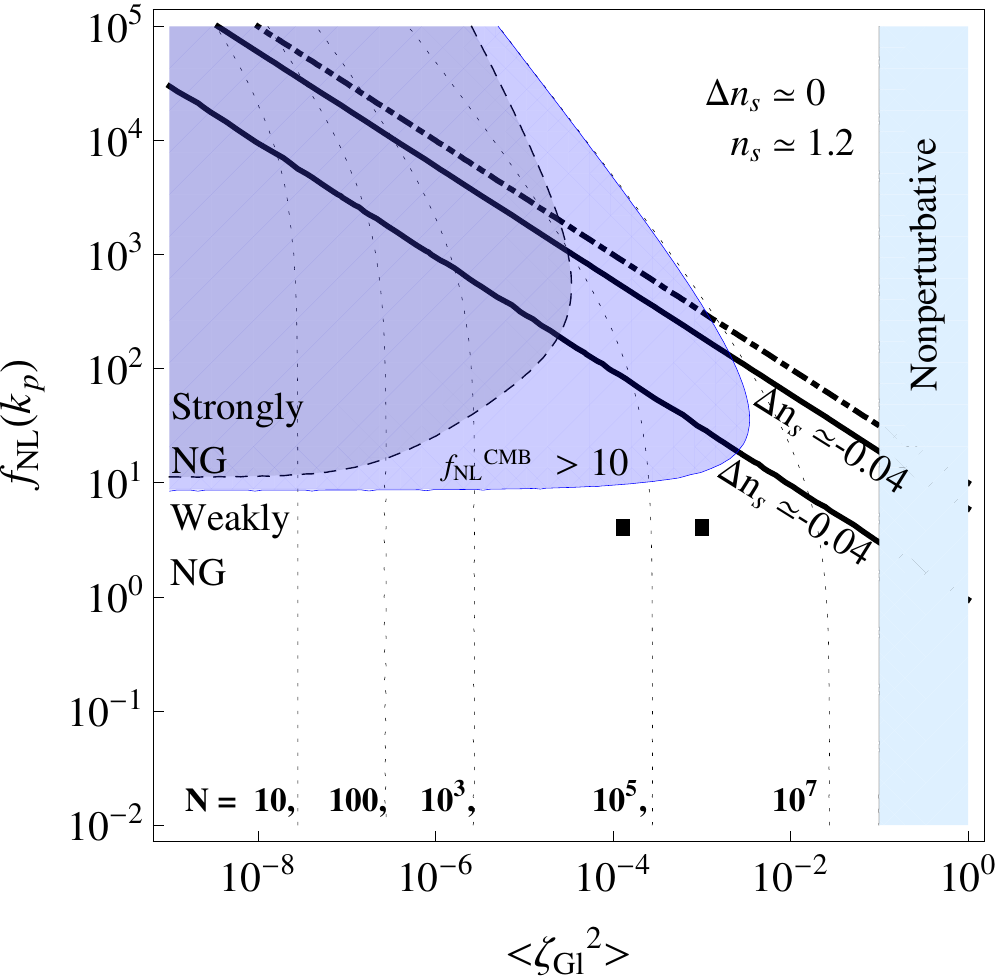} & \includegraphics[width=.49\textwidth
]{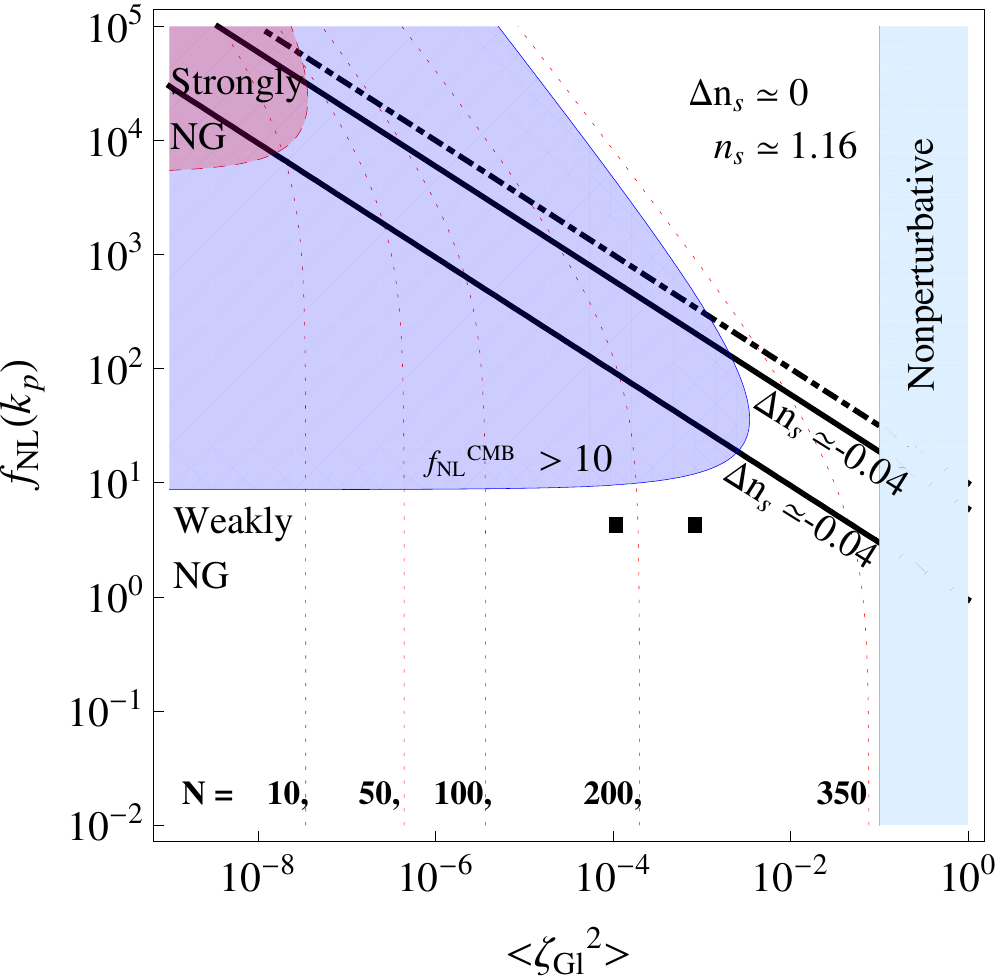} \\
$\bf{n_{\zeta}=1,~n_f=-0.1}$ &\hspace{.7 cm} $\bf{n_{\zeta}=0.96,~n_f=-0.1}$ \\
\hspace{-.5cm}\includegraphics[width=.49\textwidth]{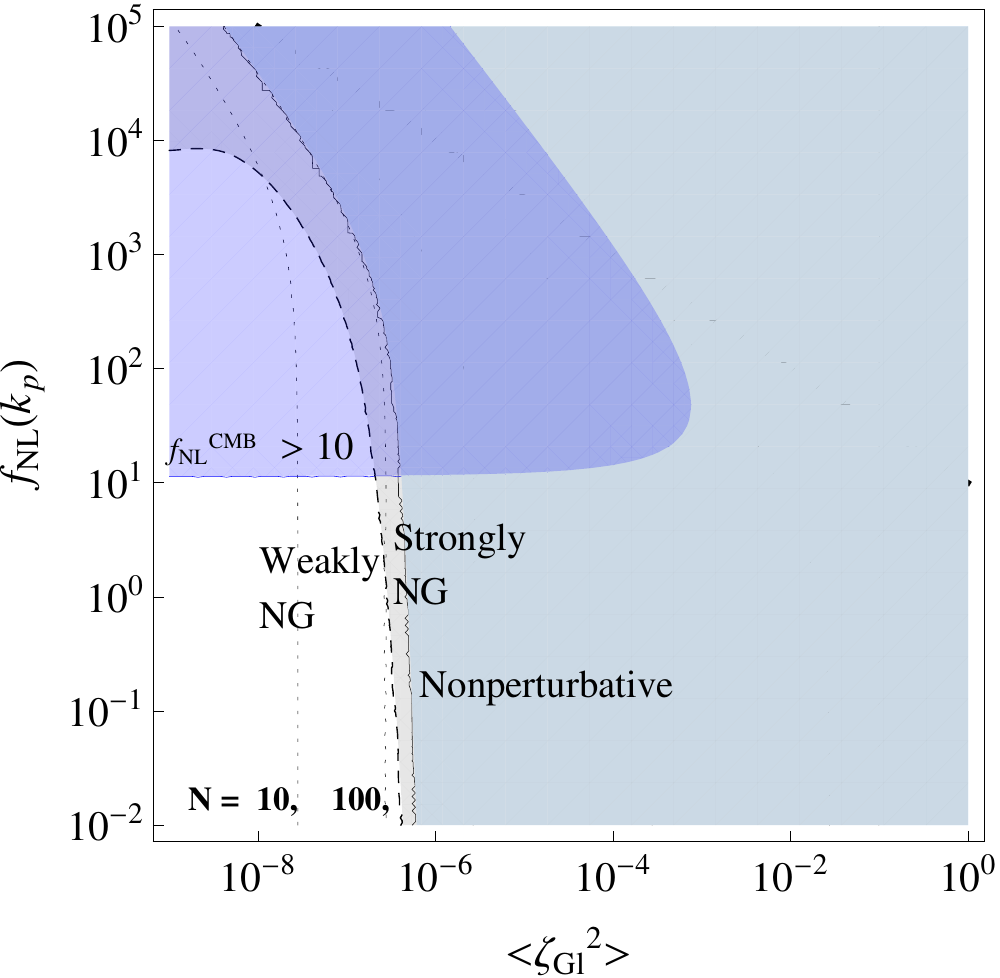} & \includegraphics[width=.49\textwidth
]{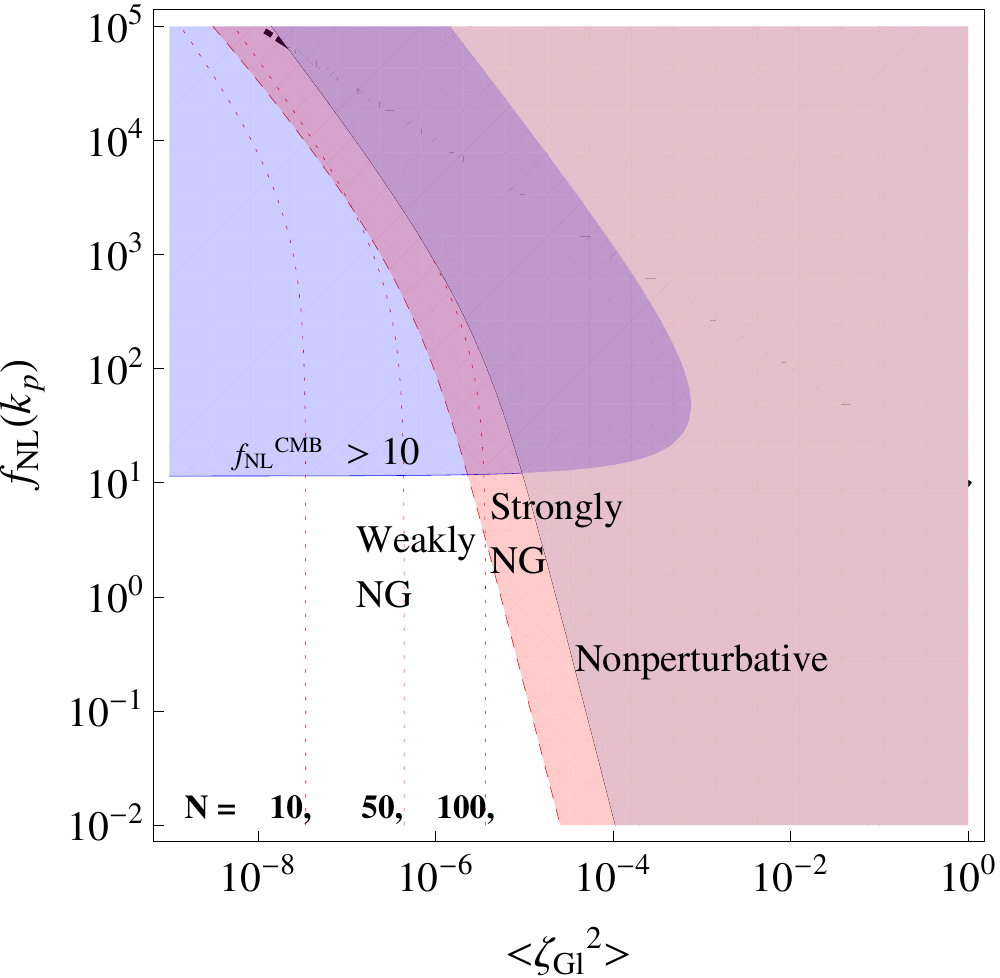} \\
\end{tabular}
\label{ConstraintsFNL}
\end{figure}

\begin{figure}
\caption{Left panel: a model with a globally flat power spectrum, but which contains subvolumes where a red tilt would be observed. Right panel: a model with global parameters naively matched to observations that nonetheless contains a significant number of subvolumes with a spectral index at odds with observations. Both cases show single-source perturbations with the running of $f_{\rm NL}$, $n_f$, plotted against the parameters controlling the size of the bias, $\frac{6}{5}f_{\rm NL}(k_p) \langle \zeta_{Gl}^2\rangle^{1/2}$. This figure assumes positive $f_{\rm NL}$ and a blue running of $f_{\rm NL}$.  The running of the power spectrum is flat ($n_s(k_p) \simeq n_\zeta = 1$) and red ($n_s(k_p) \simeq n_\zeta= 0.96$) to within $\sim0.01$ below the dotted-dashed lines in the left and right panels, respectively. Above the dotted-dashed lines the loop correction to the running of the power spectrum becomes large  ($n_s(k_p)-n_\zeta>0.01$). Dashed lines indicate regions where the non-Gaussian cumulant $\mathcal{M}_3 > 1$ for the number of superhorizon e-folds indicated. The dotted line indicates the nonperturbative region ($\langle[\frac{3}{5}f_{\rm NL} \star(\zeta_{Gl}^2-\langle\zeta_{Gl}^2\rangle)]^2 \rangle\gtrsim 0.1$) for $N > 10^3$ and $N > 100$ in the left and right panels, respectively. The grey space shows what region is excluded at $99\%$ confidence by the Planck measurement $n\o_s = 0.9603 \pm 0.0073$, assuming an underdense subsample with a $-1 \sigma$ background fluctuation.}
\centering
\begin{tabular}{cc}
\hspace{1.5 cm}$\bf n_s(k_p)\simeq \bf{n_{\zeta}=1}$ \hspace{5.0 cm} $\bf n_s(k_p)\simeq\bf{n_{\zeta}=0.96}$ \\
\includegraphics[width=.5\textwidth]{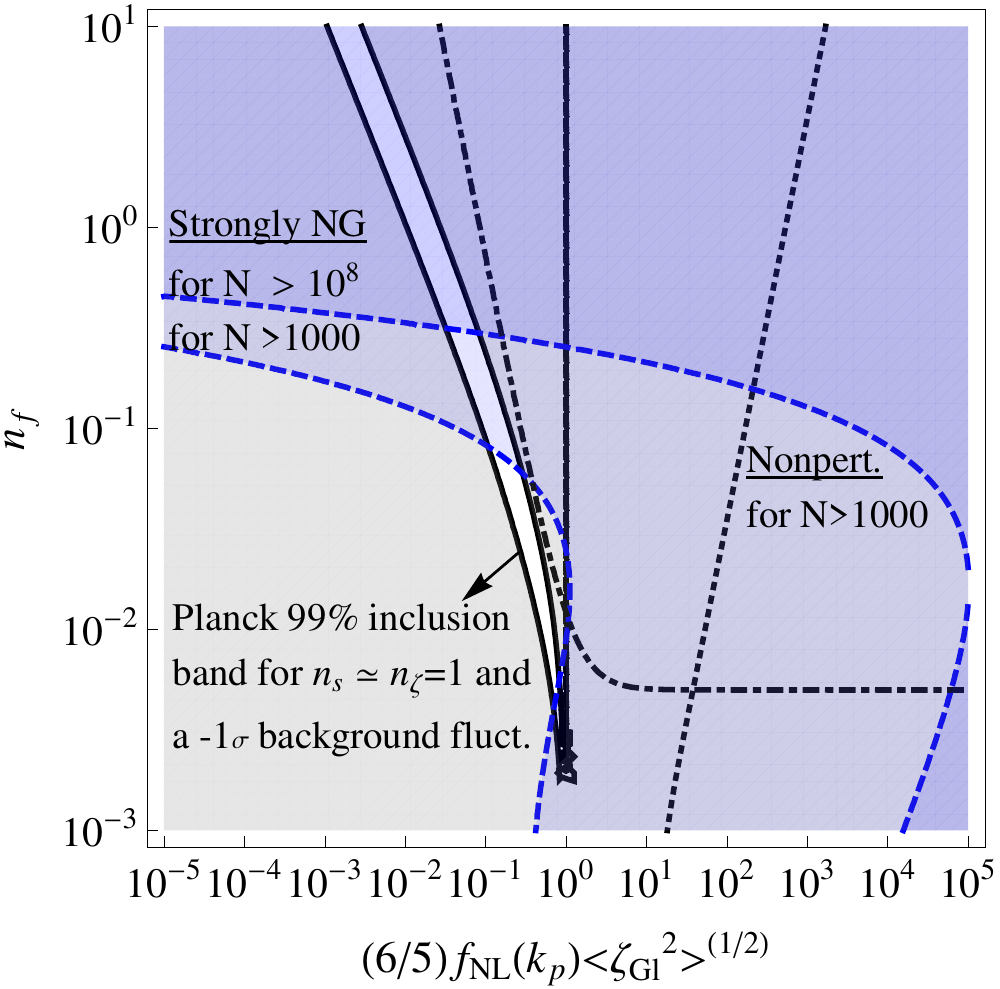}\includegraphics[width=.5\textwidth]{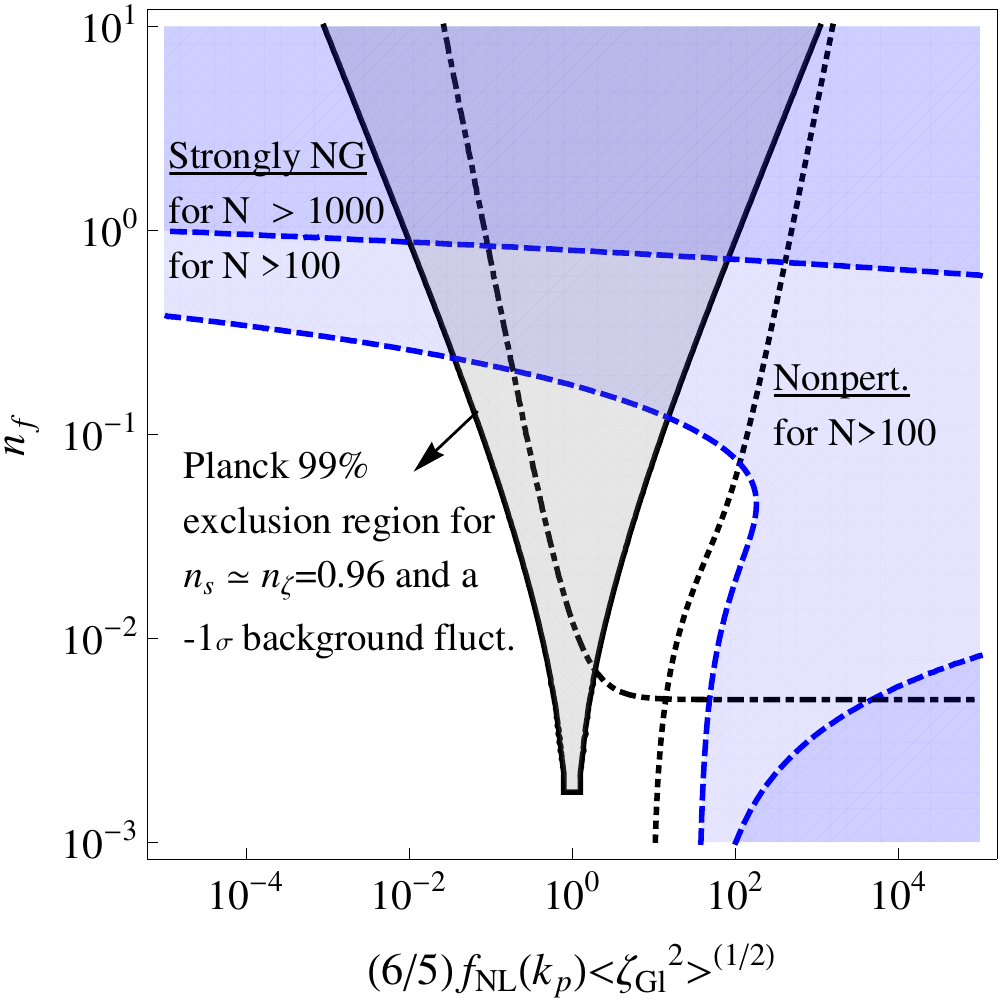}
\end{tabular}
\label{exclusionplanck}
\end{figure}

Next, consider a single source local ansatz with scale-dependent non-Gaussianity parameterized by $n_f\equiv\frac{d\ln f_{\rm NL}}{d\ln k}$. The parameter spaces for large volume statistics with $n_f=\pm0.1$ and a red or scale-invariant power spectrum $\mathcal{P}_G$ are shown in Figure~\ref{ConstraintsFNL}. All plots here assume an overdense subsample with a $+0.5 \sigma$ background fluctuation. Remarkably, the upper left panel shows that the super-Hubble universe could have a flat spectral index $n_{\zeta}=1$, while still being consistent with \textit{Planck}'s observations at the Hubble scale. Conversely, the right panels demonstrate that models with running non-Gaussianity which predict $n_{\zeta}=0.96$ over a super-Hubble volume will typically yield a range of values for $n_s\o$ on observable scales in Hubble-sized subsamples. (The spectral index $n_s(k_p)$ on observable scales is only well approximated by $n_{\zeta}$ if $\frac{6}{5}f_{\rm NL}(k_p)\langle\zeta_{Gl}^2\rangle^{1/2}$ is sufficiently small; we will see in Figure \ref{exclusionplanck} below that this is still consistent with a sizeable shift $|\Delta n_s|$.)

In these plots we require $f^{\rm CMB}_{\rm NL}<10$ for a typical background fluctuation $\zeta_{Gl}=0.5\langle\zeta_{Gl}^2\rangle^{1/2}$. Due to the dependence of $f^{\rm CMB}_{\rm NL}$ on $n_f$ this condition is slightly stronger for positive $n_f$, which can be seen by comparing the upper and lower diagrams in Figure \ref{ConstraintsFNL}. On the other hand, for larger background fluctuations, $|\zeta_{Gl}|>0.5\langle\zeta_{Gl}^2\rangle^{1/2}$, the condition $f^{\rm CMB}_{\rm NL}<10$ excludes less parameter space. 

In Figure \ref{ConstraintsFNL} we compare only two types of spectral indices, $n_{\zeta}<1$ and $n_{\zeta}=1$. While the spectral index $n_{\zeta}$ does not directly affect the parameter space constrained by $f^{\rm CMB}_{\rm NL}$ and $\zeta_G<1$, it does have the following two effects:
\begin{enumerate}
\item
A red tilt in the power spectrum gives superhorizon modes more power, and biases the subvolumes more strongly (for fixed $f_{\rm NL}(k)$). Thus, a given value of $\langle\zeta_{Gl}^2\rangle$ corresponds to a smaller (larger) number of e-folds in the case of a red tilt (blue tilt), as shown in Figure~\ref{ConstraintsScaleInv}, so it is easier to realize a large shift to a global red tilt than to a global blue tilt.
In fact, as previously noted in the discussion of Figure \ref{ConstraintsScaleInv}, imposing the requirement that $\mathcal{P}\o_{\zeta}$ be typical of subvolumes for scenarios with a blue tilt $n_{\zeta}-1$ causes $\langle\zeta_{Gl}^2\rangle$ to converge to a particular value as $N$ is increased.
\item
A red tilt in the power spectrum can relax the constraint from requiring weak global non-Gaussianity, as seen by comparing the right panels in Figure~\ref{ConstraintsFNL} to the left panels. For example, when $n_f>0$, a red tilt in the power spectrum gives more relative weight in $\mathcal{M}_3$ to the more weakly coupled superhorizon modes and damps the power of strongly coupled subhorizon modes. Note that the bottom two panels in Figure \ref{ConstraintsFNL} permit about the same number of super-horizon e-folds of weakly non-Gaussian parameter space. In the right panel the power removed from subhorizon e-folds by $n_{\zeta}<1$ is balanced by power added to superhorizon e-folds leading to a larger background $\langle \zeta_{Gl}^2 \rangle$ per e-fold permitted for perturbative statistics as compared to the bottom left panel.
\end{enumerate}
For these reasons single-source scenarios with a red power tilt in the large volume have the most significant range of cosmic variance due to subsampling.

The solid black lines in Figure \ref{ConstraintsFNL} show $\Delta n_s(k_p)=-0.04$ in subvolumes with a $+0.5\sigma$ background fluctuation ($\zeta_{Gl} =+0.5\langle \zeta_{Gl}^2 \rangle^{1/2}$), and thus show part of the parameter space where $|\Delta n_s(k_p)|$ can be observationally significant. Here we have neglected the subhorizon one-loop correction $\langle \zeta_{Gs}^2(k_p) \rangle \ll \zeta_{Gl}^2$; this breaks down for small $N$ but is valid outside of the region of parameter space excluded by the requirement $f_{\rm NL}^{\rm CMB} < 10$. Rewriting Eq. \eqref{nsprime} for a single source scenario ($\xi_m=1$),
\bea
\Delta n_s^{\rm single~ source}\simeq \frac{n_f\Big( \frac{12}{5} f_{\rm NL}\zeta_{Gl}\big(1-\frac{36}{25}f_{\rm NL}^2\langle\zeta_{Gl}^2\rangle\big) + \frac{72}{25} f_{\rm NL}^2 (\zeta_{Gl}^2-\langle\zeta_{Gl}^2\rangle) \Big)}{\big(1+\frac{6}{5}f_{\rm NL}\zeta_{Gl}\big)^2\big(1+\frac{36}{25}f_{\rm NL}^2\langle\zeta_{Gl}^2\rangle\big)}.
\label{nsprimesinglesource}
\eea
Assuming $n_f=0.1$ and $\zeta_{Gl}=0.5\langle\zeta_{Gl}^2\rangle^{1/2}$, we can solve this equation to show that $\Delta n_s^{\rm single~ source}=-0.04$ when $f_{\rm NL}(k_p)\langle\zeta_{Gl}^2\rangle^{1/2}=0.94$ or 5.9, which are the equations of the two black lines plotted in Figure \ref{ConstraintsFNL}. These lines assume positive $f_{\rm NL}$ in the large volume, $f_{\rm NL}>0$, but they remain the same for $f_{\rm NL}<0$ and a $-0.5 \sigma$ background fluctuation. For values of $|n_f|$ larger or smaller than $0.1$, the distance between these lines grows or shrinks in parameter space. Of course, for the full expression of $\Delta n_s$ and a different set of parameter choices, there can be more than two solutions of $|\Delta n_s|=0.04$. For positive $f_{\rm NL}(k_p) \zeta_{Gl}$ (see below), the typical size of $\Delta n_s$ is largest in the region between these lines ($\frac{6}{5}f_{\rm NL}(k_p) \zeta_{Gl}\sim1$) and falls towards zero on either side.

The upper dotted-dashed lines mark where $\frac{6}{5}f_{\rm NL}(k_p)\langle\zeta_{Gl}^2\rangle^{1/2}$ is large ($\mathcal{O} (10)$). When that quantity is large, $\Delta n_s\simeq\frac{-n_f}{\frac{3}{5}f_{\rm NL}\langle \zeta_{Gl}^2 \rangle^{1/2}}$ and thus approaches zero as indicated in Figure \ref{ConstraintsFNL}. Note that in this region the observed spectral index is $n_s\o \simeq n_s \simeq n_\zeta + 2n_f$, so for the parameter choices in Figure \ref{ConstraintsFNL} the \textit{Planck} satellite excludes the region above the dotted-dashed lines. All lines and contours in Figure \ref{ConstraintsFNL} assume that $\frac{6}{5}f_{\rm NL}(k_p) \zeta_{Gl} > 0$ (eg, overdense fluctuations with positive $f_{\rm NL}$). If this figure assumed $\frac{6}{5}f_{\rm NL}(k_p) \zeta_{Gl} < 0$ (eg, overdense fluctuations with negative $f_{\rm NL}$), the area in parameter space near the line $\frac{6}{5}f_{\rm NL}(k_p) \langle\zeta_{Gl}^2\rangle^{1/2}=1$ would be excluded. For further discussion of parameter space with $\frac{6}{5}f_{\rm NL} \zeta_{Gl} < 0$, see the discussion after Eq. \eqref{nsprime}.

Figure \ref{ConstraintsFNL} shows that, under the conditions we have imposed and the spectral indices considered, only scenarios where the bispectral tilt is not very red have typical subvolumes where the observed spectral index varies by an amount that is cosmologically interesting for us, $|\Delta n_s|\gtrsim 0.01$. A blue bispectral index may avoid the current observational constraints, which do not probe particularly small scales, and easily remain globally perturbative and weakly non-Gaussian (see paragraph below). In contrast, the bottom panels of Figure \ref{ConstraintsFNL} illustrate that for either spectral index, a scenario with $n_f<0$ will be nonperturbative in the interesting part of parameter space where $|\Delta n_s| \sim 0.04$. (In addition, there is only a small window with strongly non-Gaussian but perturbative global statistics.) If both the power spectrum and non-Gaussianity increase in the IR, as in the lower right panel of Figure \ref{ConstraintsFNL}, the statistics will be strongly non-Gaussian across parameter space for a small number of superhorizon e-folds.

The upper panels of Figure \ref{ConstraintsFNL} illustrate a feature discussed in Section \ref{subec:bispect}: $\frac{6}{5}f_{\rm NL}(k_p) \langle\zeta_{Gl}^2\rangle^{1/2} \gtrsim 1$ does not necessarily imply a large cumulative skewness, $\mathcal{M}_3\gtrsim1$. The dashed curves fix $\mathcal{M}_3=1$ as a function of superhorizon e-folds, which are determined at each point in parameter space by the observed level of the power spectrum along with $n_f,~f_{\rm NL}$ and $\langle \zeta_{Gl}^2 \rangle$. In regions where $\mathcal{M}_3<1$ but $f_{\rm NL}(k_p)\langle\zeta_{Gl}^2\rangle^{1/2}\gtrsim1$, there are a sufficient number of superhorizon modes with weaker coupling ($n_f>0$) damp the total non-Gaussianity. To elaborate, in the limit $n_f(N+N_{\text{sub}})\gg1$, Eq.~(\ref{M3ns1}) gives $\mathcal{M}_3\propto[\langle\zeta_{Gl}^2\rangle/N(N+N_{\rm sub})]^{1/2}$. For $f_{\rm NL}(k_p)\langle\zeta_{Gl}^2\rangle^{1/2}\ll1$, $N=\langle\zeta_{Gl}^2\rangle/\mathcal{P}\o_{\zeta}$ and so $\mathcal{M}_3$ becomes independent of $\langle\zeta_{Gl}^2\rangle$ in the limit $N\ll N_{\text{sub}}$. For $f_{\rm NL}(k_p)\langle\zeta_{Gl}^2\rangle^{1/2}\gg1$, on the other hand, $\mathcal{M}_3\propto1/f_{\rm NL}(k_p)\langle\zeta_{Gl}^2\rangle^{3/2}$, so large $f_{\rm NL}(k_p)$ and sufficiently large $\langle\zeta_{Gl}^2\rangle$ are needed to keep the total non-Gaussianity small, \textit{and} $\mathcal{P}\o_{\zeta}\sim2\times10^{-9}$ typical in subvolumes, as seen in the upper left panel of Figure \ref{ConstraintsFNL}. Note that throughout this analysis, we have assumed $n_f$ is constant for all $N_{\text{sub}}=60$ subhorizon e-folds, so that for blue $n_f$ non-Gaussianity continues to grow on subhorizon scales where nonlinear evolution has taken over. If this condition is relaxed, the conditions from weak non-Gaussianity are less restrictive.

\begin{figure}
\caption{The probability of finding a shift in the spectral index in subvolumes. Left panel: The variance plotted here corresponds to about 195 extra e-folds in a model with $n_{\zeta}=0.96$ or $4\times10^4$ extra e-folds for a scale-invariant spectrum. Right panel: The variance here is consistent with about 240 extra e-folds in a model with $n_{\zeta}=0.96$ or $5\times10^5$ extra e-folds for a scale-invariant spectrum. In both panels the solid black lines show a bispectral index of $n_f=0.05$ while the dotted blue lines show $n_f=0.1$. In the right panel about 24\% (6\%) of subvolumes in the $n_f=0.1$ ($n_f=0.05$) have $\Delta n_s\geq0.02$ and 17\% (5\%) have $\Delta n_s\leq-0.04$. The points in parameter space that correspond to the dotted lines ($n_f=0.1$) are shown with black squares in Figure \ref{ConstraintsFNL}.}
\vspace{0.5cm}
\centering
\begin{tabular}{cc}
\includegraphics[width=.5\textwidth]{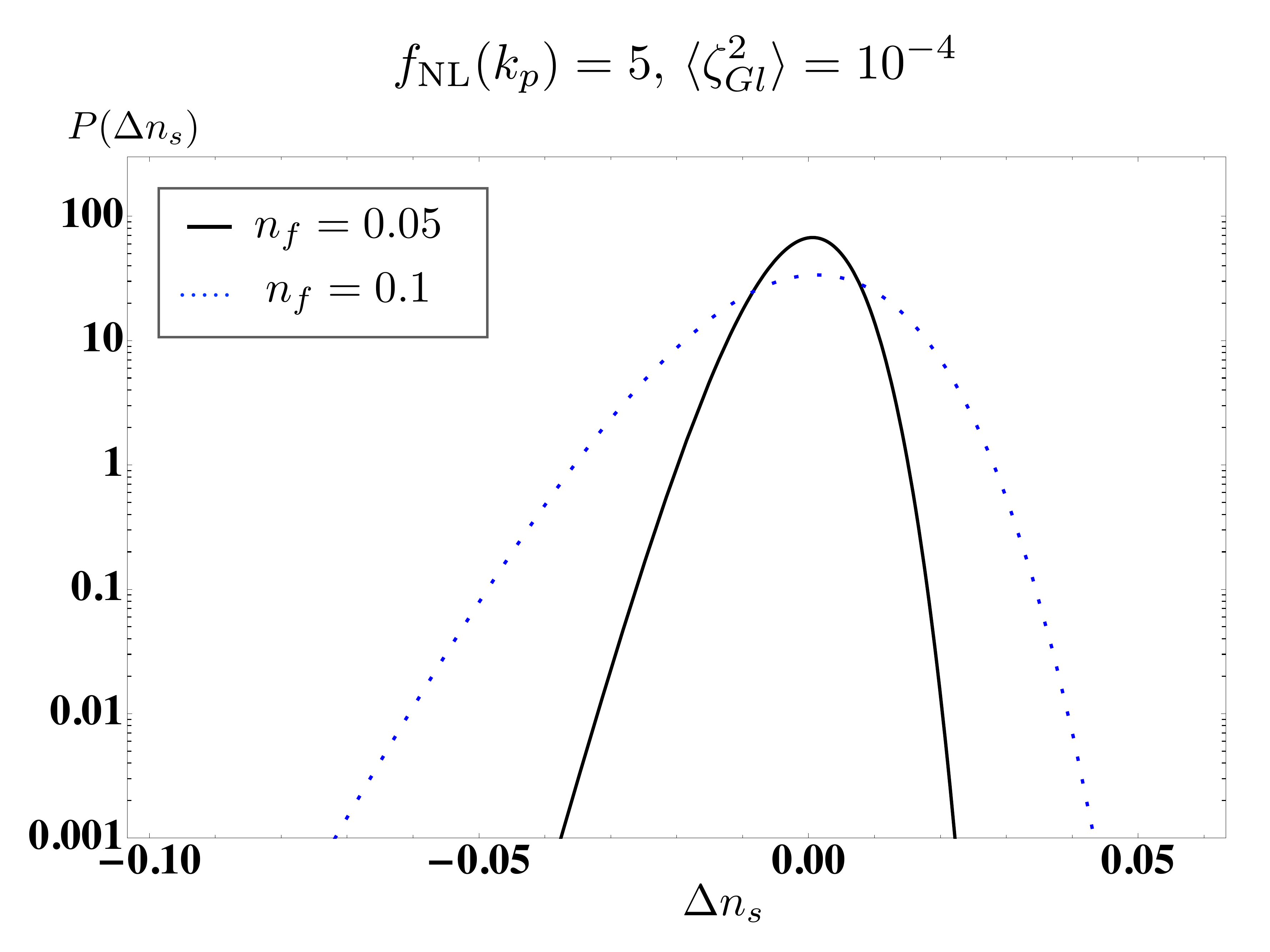}\includegraphics[width=.5\textwidth]{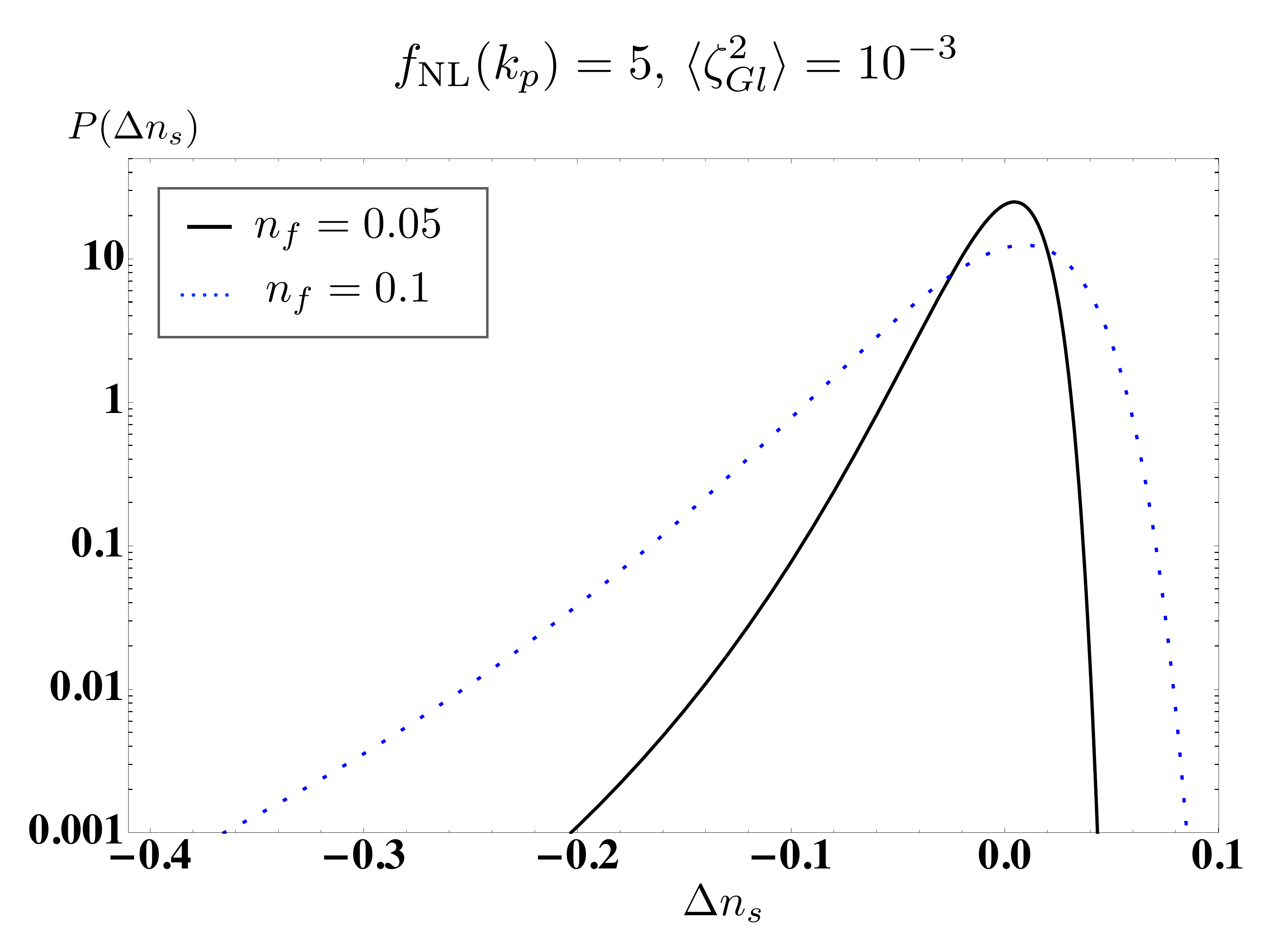}
\end{tabular}
\label{fig:ProbDeltans}
\end{figure}

Figure \ref{fig:ProbDeltans} shows the probability distribution for the shift $\Delta n_s$ for the parameters in part of the range of interest for the blue bispectral index shown in the top panels of Figure \ref{ConstraintsFNL}. Both panels show examples that (for appropriate choices of large volume parameters) give local power spectra amplitude and $f_{\rm NL}^{\rm CMB}$ consistent with our observations. Notice that the distribution on the right is substantially less Gaussian than the distribution on the left. This trend continues if one considers larger $\langle\zeta_{Gl}^2\rangle$ while keeping all other parameters fixed.

In Figure \ref{exclusionplanck} we show regions of parameter space in the $(\frac{6}{5}f_{\rm NL}(k_p) \langle \zeta_{Gl}^2\rangle^{1/2}, n_f)$ plane that are consistent with the Planck measurement $n\o_s = 0.9603 \pm 0.0073$. Assuming that the scalar power spectrum in the full volume of the mode-coupled universe is completely flat, $n_{\zeta}=1$, we see that $\frac{6}{5}f_{\rm NL}(k_p) \langle \zeta_{Gl}^2\rangle^{1/2}$ must be at least $\mathcal{O} (10^{-1})$ and for weakly non-Gaussian statistics, more than a hundred superhorizon e-folds are required. It is interesting to note that in the case of a blue-tilted $f_{\rm NL}$, a larger running non-Gaussianity $n_f$ loosens parameter constraints coming from requiring perturbative statistics $\langle [f_{\rm NL} (k) \zeta_{Gl}^2]^2 \rangle\lesssim 0.1$. Although the dotted lines in Figure \ref{exclusionplanck} will shift to the left with more superhorizon e-folds, these curves exclude less parameter space as $n_f$ becomes larger. This is because we have assumed $n_f$ is blue and constant so $f_{\rm NL}$ is driven to smaller values in the IR and $\langle [f_{\rm NL} (k) \zeta_{Gl}^2]^2 \rangle$ becomes smaller for larger $n_f$. Notice the shift in the non-perturbative line in the right panel that occurs at $n_f > |n_{\zeta}-1|$: if the running of the power spectrum is larger than the running of $f_{\rm NL}(k)$, then the running of the power spectrum will dominate the variance of  the local quadratic term over superhorizon modes, because $f_{\rm NL}^2 (k) \mathcal{P}_{G}(k)^2 \propto k^{2(n_f+n_{\zeta}-1)}$. Lastly, the right panel of Figure \ref{exclusionplanck} shows once again that for a blue tilted $f_{\rm NL}$, the weakly non-Gaussian parameter space enlarges with the number of superhorizon e-folds, because $f_{\rm NL}$ is driven to very small values over more superhorizon e-folds, decreasing the value of $\mathcal{M}_3$.

\begin{figure}
\caption{{\bf Top panel:} The contributions to the power spectrum $\mathcal{P}_{G}(k)$ and $\mathcal{P}_{\zeta}^{\text{1-loop}}(k)\simeq\frac{36}{25}f_{\rm NL}^2(k)\langle\zeta_{Gl}^2\rangle\mathcal{P}_{G}(k)$ are shown, for the following parameter choices: $n_{\zeta}=0.95$, $n_f=0.05$, $f_{\rm NL}(k_p)\langle\zeta_{Gl}^2\rangle^{1/2}=3$. The total power spectrum is shown with a thin black line, and the corresponding shifted power spectra for a subvolume with a $+0.1\sigma$ background fluctuation is shown with a thick black line. The vertical scale can be fixed so $\mathcal{P}\o_{\zeta}$ matches the observed value.
{\bf Bottom panel:} Parameter space for single source non-Gaussianity with $n_{\zeta}=0.95$ and $n_f=0.05$ is shown. The dotted-dashed line indicates $f_{\rm NL}(k_p)\langle\zeta_{Gl}^2 \rangle^{1/2}=10$, both black lines indicate $\Delta n_s =-0.065$ for a $+0.1 \sigma$ background fluctuation, and the red circle indicates the parameter space congruent with the top panel. Dotted lines show the indicated number of superhorizon e-folds for a $+0.1 \sigma$ bias. The exclusion regions are marked the same as those in Figure \ref{ConstraintsFNL}, but these assume a $+0.1 \sigma$ bias.}
\centering\begin{tabular}{ccc}
\hspace{1cm}\includegraphics[width=.6\textwidth]{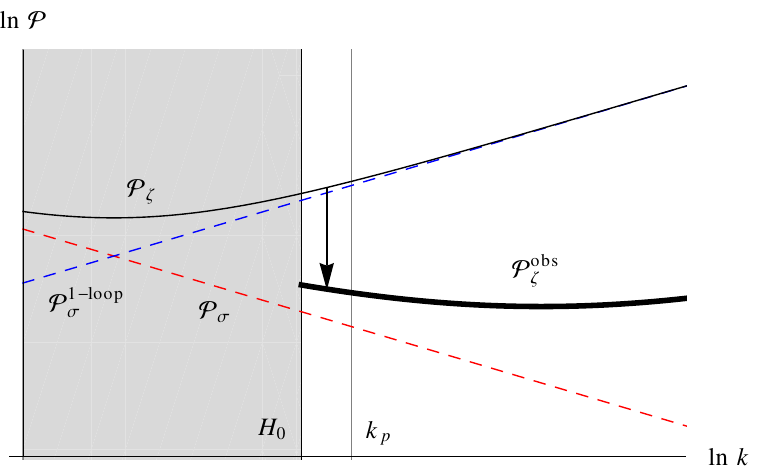} \\
$\bf{n_\zeta=0.95}$ ~ $\bf{n_f=0.05}$ \\
\includegraphics[width=.5\textwidth]{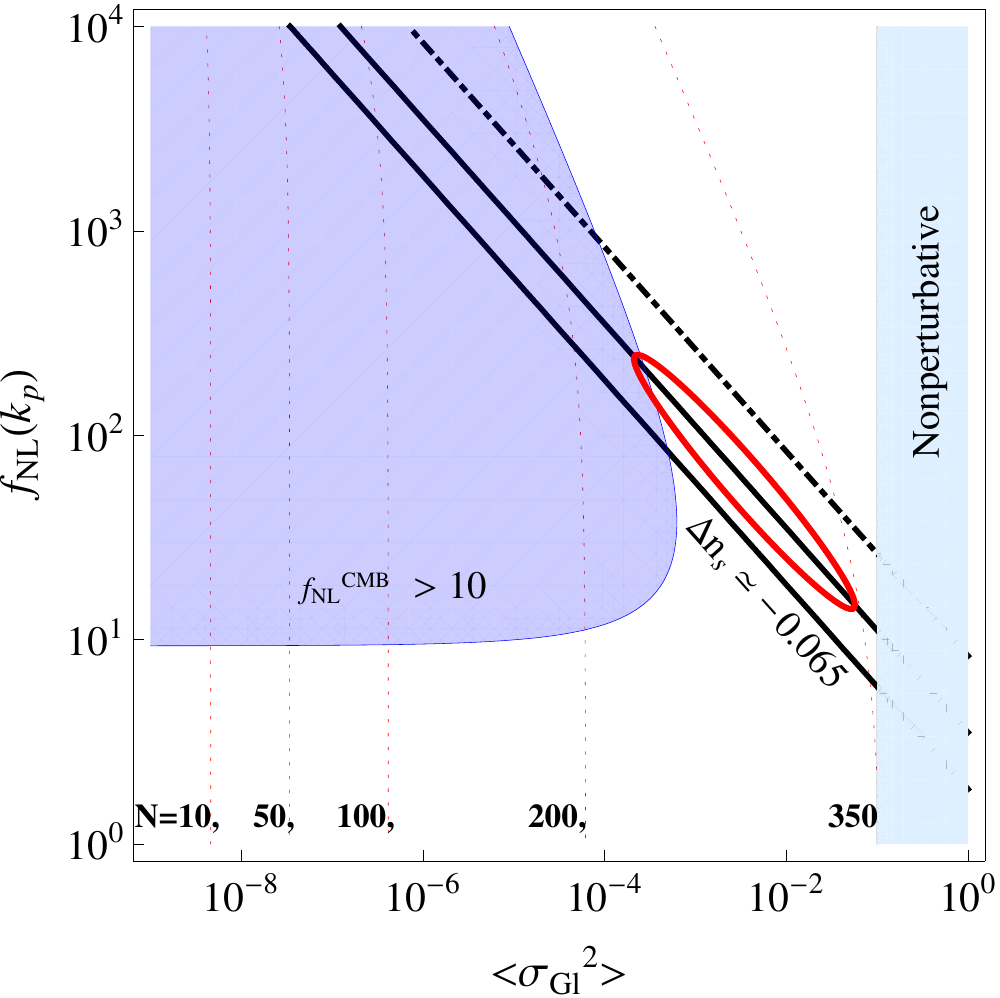}
\end{tabular}
\label{SingleSourcePowerShift}
\end{figure}

To conclude this section, Figure \ref{SingleSourcePowerShift} illustrates a single-source scenario in which a power spectrum which appears blue-tilted in the large volume on short scales can appear red on the same scales in a subvolume. On scales where $P_{\zeta}(k)\simeq P_{G}(k)$, $n_s(k)\simeq n_{\zeta}$, whereas on scales where the 1-loop contribution dominates $\mathcal{P}_{\zeta}^{\text{1-loop}}(k)\simeq\frac{36}{25}f_{\rm NL}^2(k)\langle\zeta_{Gl}^2\rangle\mathcal{P}_{G}(k)$ and the spectral index will be $n_s(k)\simeq n_{\zeta}+2n_f$. If the transition of power takes place on a scale near the observable range of scales ($f_{\rm NL}(k_p)\langle\zeta_{Gl}^2\rangle^{1/2}=\mathcal{O}(1)$), the observed spectral index can be shifted. For example, if $\zeta_{Gl}^2<\langle\zeta_{Gl}^2\rangle$, the blue-tilted $f_{\rm NL}^2\langle\zeta_{Gl}^2\rangle$ contribution loses power in the subvolume, and if $f_{\rm NL}(k_p)\zeta_{Gl}>0$, the red-tilted piece gains power (compare Eqs. \eqref{pzeta}, \eqref{Pprime1}). This scenario is shown in Figure \ref{SingleSourcePowerShift}. Note that as long as $f_{\rm NL}(k_p)$ is not extremely large (which would violate the constraint on $f_{\rm NL}^{\rm CMB}$ for the value of $f_{\rm NL}(k_p)\langle\zeta_{Gl}^2\rangle^{1/2}$ chosen here), $\zeta_{Gl}\gg\langle\zeta_{Gs}^2(k)\rangle^{1/2}$ and the $1$-loop contribution to $\mathcal{P}\o_{\zeta}$ is very small, suppressed by a factor of $\langle\zeta_{Gs}^2(k)\rangle/\zeta_{Gl}^2$.\\

\hspace{-25pt}\textbf{Example III: Multiple sources with running $\xi_m(k)$.}

In the single-source case, a large shift to the observed spectral index could only occur if the 1-loop contribution to the power spectrum dominated on small scales. With two sources, a significant shift to $n_s$ can be consistent with weak non-Gaussianity $\xi_m(k)f_{\rm NL}(k)\langle\sigma_{Gl}\rangle^{1/2}<1$ on all scales. If the running of the 1-loop contribution lies between the runnings $n_\sigma \equiv \frac{d \ln \mathcal{P}_\sigma (k)}{d \ln (k)}$ and $n_{\phi} \equiv \frac{d \ln \mathcal{P}_\phi (k)}{d \ln (k)}$ of the Gaussian contributions to the total power, then it will be subdominant on large and small scales.

The transition of power between $\sigma_G$ and $\phi_G$ takes place over a finite range of scales, over which $n_s$ changes from $n_{\sigma}$ to $n_{\phi}$. If the power spectrum of $\phi_G$ is blue and dominates on small scales ($\xi_m(k\gtrsim H_0)\ll1$), and the Gaussian contribution from $\sigma$ is red and dominates on large scales ($\xi_m(k<<H_0)\simeq 1$), then the background $\zeta_l\simeq\sigma_l$ for any subvolume couples to and biases the local statistics. For example, a globally flat or blue spectral index $n_s(k>H_0)>1$ can again appear red, $n\o_s<1$, in a subvolume. The shift to $n_s$ can come only from the modulation of power in $\sigma$ relative to $\phi_G$, and need not rely on running non-Gaussianity $n_f\neq0$. That is, a large running of the difference in power of the fields can be achieved without a large level of running non-Gaussianity. This becomes apparent upon inspecting the running of $\xi_m$,
\bea
n_f^{(m)} (k)\equiv \frac{d \ln \xi_m(k)}{d \ln k} = (1-\xi_m(k))\Big[n_\sigma-n_\phi + \frac{2n_f\frac{36}{25}f_{\rm NL}^2(k) \langle \sigma_{G}^2(k) \rangle}{1+\frac{36}{25}f_{\rm NL}^2(k) \langle \sigma_{G}^2(k) \rangle}\Big].
\eea 

If $\phi_G$ is more red-tilted than $\sigma_G$, the background is uncorrelated with short-wavelength modes because $\phi_G$ dominates on large scales, $\zeta_l\simeq\phi_{Gl}$, so local statistics are not biased. Thus, both $n_{\sigma}\leq1$ and $n_{\phi}>n_{\sigma}$ are needed for a significant bias. 
\begin{figure}
\caption{Multifield parameter space for $\xi_m(k_p)=0.1$, $n_{\sigma}=0.93$, $n_{\phi}=1.005$, $n_f=0.001$. The black lines show $\Delta n_s\simeq -0.03$ for a $+3\sigma$ background fluctuation. The dotted-dashed line shows $f_{\rm NL}(k_p)\langle\sigma_{Gl}^2\rangle^{1/2}=10$. The upper left region shows the \textit{Planck} constraint on $f_{\rm NL}^{\rm CMB}$ for a $+3\sigma$ background.}
\centering
\hspace{-1cm}\includegraphics[width=.5\textwidth]{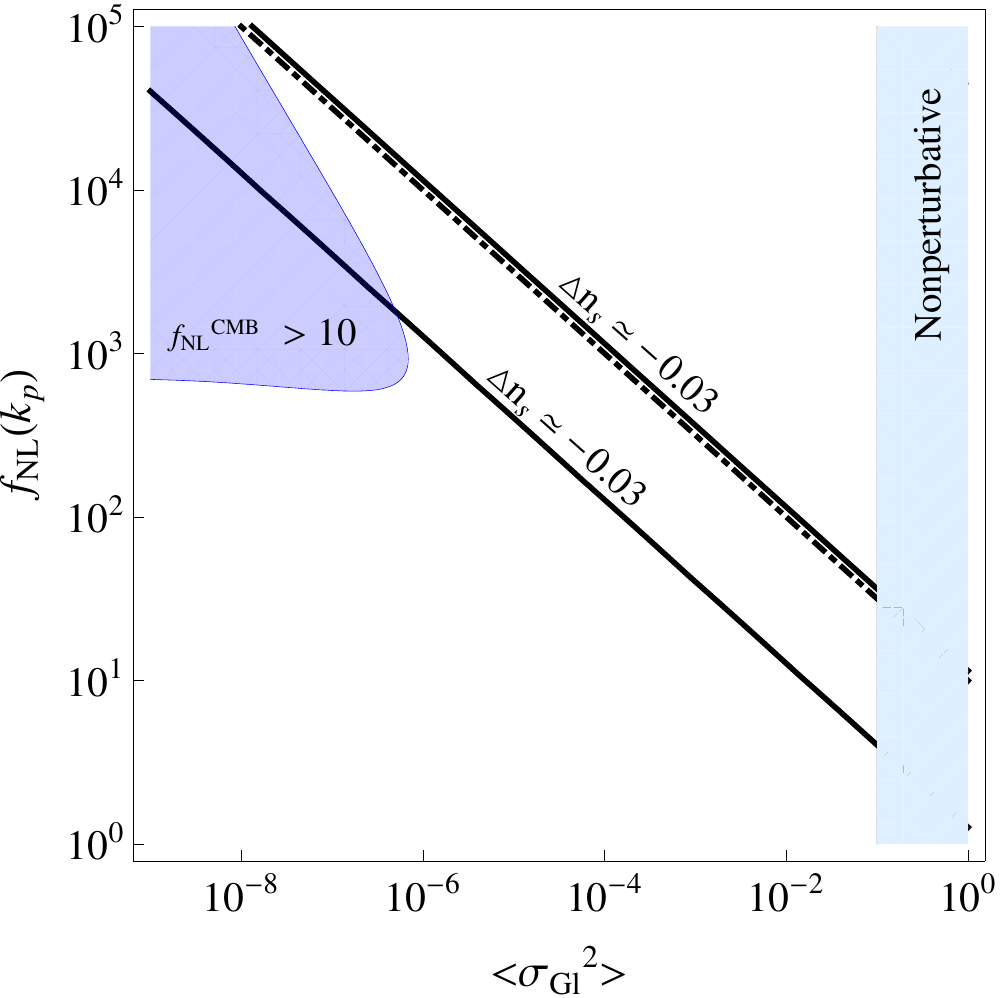}
\label{ConstraintsXi}
\end{figure}
In Figure~\ref{ConstraintsXi} we show the parameter space for the two-source scenario described above, with {$n_{\sigma}(k_p)=0.93$, $n_{\phi}(k_p)=1.005$, and $\xi_m(k_p)=0.1$. We also fix $n_f=0.001$ so that mode coupling is weaker on superhorizon scales.
As before, the upper left region shows where $f\o_{\rm NL}\gtrsim10$ in typical subvolumes. We see that adding the second source relaxes the constraint on $f_{\rm NL}$ in the $f_{\rm NL}\langle\sigma_{Gl}^2\rangle^{1/2}\ll1$ regime. This makes it possible to achieve a large shift $\Delta n_s$ for smaller values of $\langle\sigma_{Gl}^2\rangle$ and thus fewer superhorizon e-folds.

The condition $\xi_m(k_p)=0.1$ makes the field $\phi_G$ dominant on \textit{Planck} scales, so from the perspective of the large volume, the power spectrum has a blue tilt $n_s(k_p)\simeq n_{\phi}=1.005$ on scale $k_p$. However, for significant biasing ($3 \sigma$) and a small (or zero) non-Gaussian running of the coupled field $n_f = 0.001$, the black lines in Figure \ref{ConstraintsXi} denote where $\Delta n_s = -0.03$, which would be consistent with \textit{Planck} observations. Here the shift in $\Delta n_s$ is coming not from $n_f$ but from the difference in running of $P_{\sigma,NG}$ and $P_\phi$, $n_f^{(m)}$, as the red-tilted $P_{\sigma,NG}$ is amplified due to the strong background overdensity. It is also interesting to note that a cursory survey of background fluctuations reveals that biases less than $|3 \sigma|$ yield no $\Delta n_s$ corrections smaller than $-0.03$, which would seem to partly exclude these parameters for typical Hubble-sized subsamples. In the limit of very small $\xi_m(k_p)$, $\phi_G$ dominates the power and scale-dependence on observable scales, so unless the bias is extremely strong, any shift in the power and scale-dependence from the $\sigma$ field will be too small to affect $n\o_s$.

\hspace{-25pt}\textbf{Summary.}

In summary, a significant shift to the observed spectral index from correlations with long-wavelength background modes is possible under the following conditions:
\begin{enumerate}
\item \textit{A red tilt for the field with mode coupling}, $n_{\sigma}\leq1$ ($n_{\zeta}\leq1$ in the single-source case), is necessary for the cumulative power $\langle\sigma_{Gl}^2\rangle$ on superhorizon scales to be large enough to significantly bias local statistics.
\item \textit{A blue bispectral index $n_f\geq0$ for $f_{\rm NL}(k)$} (assuming constant $n_f$) is needed to remove the power from the non-Gaussian term on large scales so that strong coupling of short-scale modes to background modes is consistent with weak global non-Gaussianity and $\zeta$ being perturbative, while having enough background modes to give a large bias.
\item \textit{In a two-source scenario, the ratio of power in the non-Gaussian field to total power should have a red spectrum} ($n_f^{(m)}(k_p)\leq0$) so that the non-Gaussian field $\sigma_G$ grows relative to $\phi_G$ on large scales, causing the background $\zeta_l$ to be sufficiently correlated with local statistics. If $\phi_G$ contributes on observable scales ($\xi_m(k_p)<1$), larger values of $f_{\rm NL}(k_p)$ are consistent with observational constraints on non-Gaussianity, so a smaller background $\sigma_{Gl}$ is needed to give the same shift to $n\o_s$.
\end{enumerate}

Introducing scale-dependence into the spectral indices would relax the conditions for large $|\Delta n_s|$. Although the scenario becomes more complicated in this case, the qualitative features remain valid: scale-dependence of power spectra and non-Gaussian parameters must allow for sufficient cumulative superhorizon power that a large background $\sigma_{Gl}$ from the source with mode coupling is typical.

We note that for given large-volume statistics, the observed red tilt may not be equally consistent with a local overdensity or underdensity in $\sigma_G$. In the single-source case with $n_f>0$, for example, an overdensity (underdensity) corresponds to an increase (decrease) of power on small scales. Thus, for a scale-invariant power spectrum in the large volume, the observed red tilt $n\o_s\simeq 0.96$ could be accounted for in terms of a blue-tilted global bispectrum and local underdensity. However, without information about the global power spectrum, it would be difficult to infer whether we sit on a local underdensity or overdensity.

\subsection{The shift to the scale dependence of the bispectrum}
\label{sec:bispec}
The bispectrum may also be shifted by mode coupling coming from the soft limits of the large-volume trispectrum and from any non-Gaussian shifts to power spectrum. We can define a spectral index for the squeezed limit of the bispectrum {\it within} any particular volume as
\be
n_{\rm sq.}\equiv\frac{d\ln B_{\zeta}(k_L,k_S,k_S)}{d\ln k_L}-(n_s-1) \label{nBbigbox}
\ee
where $k_L$ and $k_S$ are long wavelength and short wavelength modes, respectively. The small volume quantity, $n\o_{\rm sq.}$, should be calculated using the observed bispectrum and the observed spectral index. For a single source, scale-invariant local ansatz, $n_{\rm sq.}=-3$. For the single source, weakly non-Gaussian, scale-dependent scenario with $g_{\rm NL}$ absent, the shift in this bispectral index between the large volume and what is observed in the small volume is
\bea
{\rm Single\; Source:}\;\;\;\;\Delta n_{\rm sq.}(k)&\equiv& n\o_{\rm sq.}(k)-n^{\rm Large Vol.}_{\rm sq.}(k)\\\nonumber
&\approx&-\frac{\frac{6}{5}f_{\rm NL}(k_L)\sigma_{Gl}\,n_f}{1+\frac{6}{5}f_{\rm NL}(k_L)\sigma_{Gl}}\;.
\label{nBsmallbox}
\eea
If $\frac{6}{5}f_{\rm NL}(k_L)\sigma_{Gl}=\frac{6}{5}f_{\rm NL}(k_L)\langle\zeta_G^2\rangle^{1/2}B\ll1$, then $\Delta n_{\rm sq.}(k)\approx-\frac{6}{5}f_{\rm NL}(k_L)\langle\zeta_G^2\rangle^{1/2}B\,n_f$. This shift is less than one in magnitude, but still relevant for interpreting bispectral indices of order slow-roll parameters.

In the two source case, there can be additional scale dependence coming from the ratio of power of the two fields. Considering only the weak coupling case, $\frac{6}{5}f_{\rm NL}(k)\sigma_{Gl}\ll1$ (and again setting $g_{\rm NL}=0$ for simplicity),
\bea
{\rm Two\; Source:}\;\;\;\;\Delta n_{\rm sq.}(k)&=&\frac{\frac{12}{5}f_{\rm NL}(k)\sigma_{Gl}}{1+\frac{12}{5}f_{\rm NL}(k)\sigma_{Gl}}n_f-\frac{\frac{6}{5}f_{\rm NL}(k)\sigma_{Gl}}{1+\frac{6}{5}f_{\rm NL}(k)\sigma_{Gl}}n_f\\\nonumber
&&-\frac{\frac{12}{5}\xi_m(k)f_{\rm NL}(k)\sigma_{Gl}}{1+\frac{12}{5}\xi_m(k)f_{\rm NL}(k)\sigma_{Gl}}(n_f+n_f^{(m)})\\\nonumber
&\approx& \frac{6}{5}f_{\rm NL}(k)\sigma_{Gl}\,n_f-\frac{12}{5}\xi_m(k)f_{\rm NL}(k)\sigma_{Gl}(n_f+n_f^{(m)})\;.
\eea
Reintroducing $g_{\rm NL}$ and higher terms would lead to additional terms, introducing scale-dependence even if $f_{\rm NL}$ in the large volume is a constant.

\subsection{Generalized local ansatz and single source vs. multi source effects}
\label{sec:generalizedlocalansatz}

The two source, weakly scale dependent local ansatz in Eq.~\eqref{eq:GenLocalRealSpace} is representative of the properties of inflation models that generate local type non-Gaussianity. For example, the scale-dependence $f_{\rm NL}(k)$ can come from curvaton models with self-interactions \cite{Byrnes:2010xd,Huang:2010cy}. The function $\xi_m(k)$ comes from the difference in power spectrum of two fields (eg, the inflaton and the curvaton) contributing to the curvature fluctuations. In typical multi-field models, the bispectral indices $n_f$, $n_f^{(m)}$ are of order slow-roll parameters (like the scale dependence of the power spectrum), and are often not constant. Generic expressions for the squeezed limit behavior of a multi-field bispectrum are given in \cite{Dias:2013rla}. The scale-dependent functions $f_{\rm NL}(k)$ and $\xi_m(k)$ are observationally relevant for tests for primordial non-Gaussianity using the bias of dark matter halos and their luminous tracers (eg. quasars or luminous red galaxies). The power law dependence of the squeezed limit on the long wavelength, small momentum mode ($n_{\rm sq.}$ from Eq.~(\ref{nBbigbox})) generates the scale-dependence of the non-Gaussian term in the bias. The dependence on the short wavelength modes generates a dependence of the non-Gaussian bias on the mass of the tracer (which is absent in the usual local ansatz). In principle, if local non-Gaussianity is ever detected, it may be within the power of future large scale structure surveys to detect some amplitude of running \cite{Shandera:2010ei}.

However, as demonstrated above, the same shape of bispectrum can be generated locally by a single source for the curvature perturbations, so the presence of the non-trivial function $\xi_m$ in the observed bispectrum does not necessarily indicate that two fundamental fields contributed to the primordial curvature perturbations. On the other hand, the presence of one Gaussian source and one non-Gaussian source for the local curvature perturbations is in principle detectable by comparing power spectra that are sensitive in different ways to the total curvature field and to just the non-Gaussian part \cite{Tseliakhovich:2010kf}. Eq.~(\ref{Bprime}) shows that in a single source scenario the local background $\sigma_{Gl}$ can act as a second field to generate the full, multi-source shaped bispectrum, but $\sigma_{Gl}$ is constant within a single volume. This `second field' does not have fluctuations on all scales, but its variations are relevant for considering a collection of subvolumes of a particular size.

\section{Mode coupling effects from a non-local factorizable bispectrum}
\label{sec:NonLocalBispect}
We have considered the effect of superhorizon modes only for the case of nearly local non-Gaussianity, but inflationary theory has generated an expanding space of models exhibiting different types of mode coupling. Intuitively, any scenario that does not couple modes of sufficiently different wavelengths should not lead to correlation functions whose amplitudes or shapes change under subsampling. As a first step towards considering the observational consequences of subsampling general non-Gaussian scenarios, it is straighforward to find corrections from the background to small-volume quantities in the case of a factorizable quadratic kernel in Fourier space with power-law dependence.

Consider a curvature perturbation in the large volume given by
\be
\zeta_{\mathbf{k}}=\phi_{G,\mathbf{k}}+\sigma_{G,\mathbf{k}}+\int_{L^{-1}}\frac{d^3p_1}{(2\pi)^3}\frac{d^3p_2}{(2\pi)^3}(2\pi)^3\delta^3(\mathbf{p}_1+\mathbf{p}_2-\mathbf{k}) F(p_1,p_2,k)\sigma_{G,\mathbf{p_1}}\sigma_{G,\mathbf{p_2}}+...,
\label{nonlocalzeta}
\ee
where
\be
F(k_1,k_2,k_3)=\sum_ja_{{\rm NL},j}(k_p)\left(\frac{k_1}{k_p}\right)^{m_{1,j}}\left(\frac{k_2}{k_p}\right)^{m_{2,j}}\left(\frac{k_3}{k_p}\right)^{m_{3,j}}
\label{kernel}
\ee
is a sum of factorizable terms with power law dependence on the momenta. On the right hand side the $a_j$ are amplitudes defined at a pivot scale $k_p$. When $\sum_i m_{i,j}\simeq0$ for every term $j$, the bispectrum is approximately scale-invariant. The kernel $F(k_1,k_2,k_3)$ can be chosen to generate a desired bispectrum with well behaved one-loop corrections to the power spectrum \cite{Scoccimarro:2011pz}.

Splitting the modes into long and short, the locally defined short wavelength modes with shifts induced from coupling to long wavelength modes from one term in the series above are
\bea
\zeta_{\mathbf{k}_s}&=&\phi_{G,\mathbf{k}_s}+\sigma_{G,\mathbf{k}_s}+\sigma_{G,\mathbf{k}_s}a_{\rm NL}(k_p)\left[\left(\frac{k}{k_p}\right)^{m_1+m_3}\sigma_{Gl}^{(m_2)}+\left(\frac{k}{k_p}\right)^{m_2+m_3}\sigma_{Gl}^{(m_1)}\right]\\\nonumber
&&+a_{\rm NL}(k_p)\int_{M^{-1}}\frac{d^3p}{(2\pi)^3}\sigma_G(p)\sigma_G(|\mathbf{k}-\mathbf{p}|)\left(\frac{k}{k_p}\right)^{m_3}\left(\frac{|\mathbf{k}-\mathbf{p}|}{k_p}\right)^{m_2}\left(\frac{p}{k_p}\right)^{m_1}
\eea
where
\be
\sigma_{Gl}^{(m_L)}\equiv\int_{L^{-1}}^{M^{-1}}\frac{d^3p}{(2\pi)^3}\sigma_{\mathbf{p}}\left(\frac{p}{k_p}\right)^{m_L}\label{m_L}\;.
\ee
When the local field is weakly non-Gaussian, the second line is small and we can rewrite the first line as
\bea
\label{nonlocalzetas}
\zeta_{\mathbf{k}_s}&\approx&\phi_{G,\mathbf{k}_s}+\sigma_{G,\mathbf{k}_s}[1+\Delta\sigma(k)]\\\nonumber
\Delta\sigma(k)&=&a_{\rm NL}(k_p)\left[\left(\frac{k}{k_p}\right)^{m_1+m_3}\sigma_{Gl}^{(m_2)}+\left(\frac{k}{k_p}\right)^{m_2+m_3}\sigma_{Gl}^{(m_1)}\right]\;.
\eea
The leading shift to the power spectrum $P\o_{\zeta}$ in a subvolume from unobservable infrared modes in one term of the series above (and assuming weak non-Gaussianity) is:
\bea\label{Pobsnonlocal}
P\o_{\zeta}(k)&=&P_{\zeta}(k)\left\{1+\xi_m(k)\left[2\Delta\sigma(k)+\Delta\sigma(k)^2\right]\right\};
\eea
where $\xi_m(k)$ is still the ratio of power in the non-Gaussian source to the total power, defined in Eq.(\ref{eq:Definexi}). In the two-field case with weak non-Gaussianity on all scales, the observed ratio of power in the two fields is related to the same ratio in the large volume by
\bea
\xi\o_m(k)&=&\xi_m(k)\,\frac{[1+\Delta\sigma(k)]^2}{1+\xi_m(k)[2\Delta\sigma(k)+\Delta\sigma(k)^2]}\;.
\eea
The induced shift to the spectral index has two terms, but assuming that, say, the first term in the square brackets in $\Delta\sigma$ is dominant and defining $m_S=m_1+m_3$, $m_2=m_L$, and $a_{\rm NL}(k)=a_{\rm NL}(k_p)(k/k_p)^{m_S}$ it is
\be
\Delta n_s(k) \approx  2a\xi_m(k)a_{\rm NL}(k)\sigma_{Gl}^{(m_L)}(n_f^{(m)}+m_S)\;.
 \ee

The bispectrum in the large volume is
\bea
B_{\zeta}(k_1,k_2,k_3)&=&a_{\rm NL}(k_p)\left(\frac{k_3}{k_p}\right)^{m_3}P_{\zeta}(k_1)\xi_m(k_1)P_{\zeta}(k_2)\xi_m(k_2)\\\nonumber
&&\times\left[\left(\frac{k_1}{k_p}\right)^{m_1}\left(\frac{k_2}{k_p}\right)^{m_2}+\left(\frac{k_1}{k_p}\right)^{m_2}\left(\frac{k_2}{k_p}\right)^{m_1}\right] +2\,{\rm perm.}
\eea
while the observed bispectrum is
\bea
B\o_{\zeta}(k_1,k_2,k_3)&=&a_{\rm NL}(k_p)\left(\frac{k_3}{k_p}\right)^{m_3}\left[\frac{P\o_{\zeta}(k_1)\xi\o_m(k_1)}{1+\Delta\sigma(k_1)}\right]\left[\frac{P\o_{\zeta}(k_2)\xi\o_m(k_2)}{1+\Delta\sigma(k_2)}\right]\\\nonumber
&&\times\left[\left(\frac{k_1}{k_p}\right)^{m_1}\left(\frac{k_2}{k_p}\right)^{m_2}+\left(\frac{k_1}{k_p}\right)^{m_2}\left(\frac{k_2}{k_p}\right)^{m_1}\right] +2\,{\rm perm.}
\eea
Consider $k_1=k_L\ll k_2\approx k_3$. If $m_2<m_1$ (so the second term in the second line of the equation above dominates), and $m_S\equiv m_1+m_3$, then in the squeezed limit the large volume bispectrum has
\be
n_{\rm sq.}(k)=-3+n_f^{(m)}+m_2\;.
\ee
The shift to the observed running of the squeezed-limit bispectrum is
\be
\Delta n_{\rm sq.}(k)=\Delta n^{(m)}_f(k)-\frac{\Delta\sigma(k)}{1+\Delta\sigma(k)}m_S \approx
\Delta\sigma\,m_S-2\xi_m\Delta\sigma(n_f^{(m)}+m_S) \;.
\ee

In the case of the generalized, two source local ansatz considered in Sections \ref{sec:Generalizing} and \ref{sec:bispec},  $a_{\rm NL}(k)=\frac{3}{5}f_{\rm NL}(k)$, $m_3=n_f$, and $m_1=m_2=0$ so $m_S=n_f$, and both terms in the square brackets of $\Delta\sigma$, Eq.(\ref{nonlocalzetas}) contribute equally, so we recover the weakly non-Gaussian limits of Eqs. \eqref{Pprime}, \eqref{nsprime}, and Eq.~\eqref{nBsmallbox}.

As a second example, consider single field inflation (with a Bunch-Davies vacuum and inflation proceeding along the attractor solution). In this case, the squeezed limit of the bispectrum diverges with the long wavelength mode no more strongly than \cite{Maldacena:2002vr,Creminelli:2004yq,Tanaka:2011aj, Pajer:2013ana},
\be
B_{\zeta}(k_L,k_S,k_S)\propto\mathcal{O}\Big(\frac{k_L}{k_S}\Big)^2P_{\zeta}(k_L)P_{\zeta}(k_S).
\ee
A bispectrum with this squeezed limit can be obtained by using the equilateral template \cite{Creminelli:2005hu} to generate a kernel $F(p_1,p_2,k)\propto-3-2p_1p_2/k^2+2(p_1+p_2)/k+(p_1^2+p_2^2)/k^2$ \cite{Scoccimarro:2011pz}. This yields a squeezed-limit bispectrum with $n_{\rm sq.}=-1$ and $m_L=2$ in Eq.(\ref{m_L}). That is, this bispectrum generates a bias $B\propto\nabla^2\zeta_{Gl}$, so there is no sensitivity of locally measured quantities to long wavelength, nearly constant modes. In single field inflation, there is a direct map between local observables and the parameters of the inflationary Lagrangian.

Finally, suppose modes are coupled through a bispectrum with a very strong squeezed-limit (eg, $n_{\rm sq.}=-4$ and $m_L=-1$). Then the biasing of local statistics may come predominantly from background modes farthest in the infrared, which are shared by many neighboring subvolumes. In other words, the dependence of the global bispectrum on the long wavelength mode is related to the average spatial gradient of the bias in the large volume.

\section{Tensor mode running as a test of inflation?}
\label{sec:tensors}
If the scale dependence of the tensor power spectrum, $n_t\equiv\frac{d\,{\rm ln}\mathcal{P}_t}{d\,{\rm ln}k}$, can someday be measured, a red tilt would be (nearly) definitive evidence for inflation and against a contracting or ekpyrotic scenario (an interesting special case is `solid inflation' \cite{Endlich:2012pz}). Would it be possible to induce a blue tilt $n_t>0$ in a subvolume the size of the observable universe when the larger volume exhibits a more typical red tilt? If so, a measurement of $n_t>0$ would not necessarily rule out standard scalar field models of inflation. Conversely, if a red tilt $n_t<0$ can be induced in a large fraction of subvolumes from non-Gaussianity in a contracting universe scenario, a measurement of $n_t<0$ may not be a smoking gun for inflation .

Consider a three-point interaction
\be
\langle\chi_{\mathbf{k}_1}\gamma^{s_1}_{\mathbf{k}_2}\gamma^{s_2}_{\mathbf{k}_3}\rangle\equiv(2\pi)^3\delta^3(\sum\mathbf{k}_i)B(k_1,k_2,k_3)\delta_{s_1s_2}
\ee
between two tensor modes $\gamma_{\bf k_i}$ with polarizations $s_i$ and one mode from a field $\chi$ (here, a scalar field for example). In the squeezed limit, this three-point function will induce a dependence of the local tensor power spectrum on superhorizon $\chi$ modes. Any choice of the Fourier space kernel that gives the correct squeezed limit of the bispectrum should show the correct shift to the local power spectrum. So, with a simple choice we find that the tensor power spectrum is shifted by the correlation with long wavelength modes $p$ as
\be
\gamma^{s_i}_{\mathbf{k}}=\gamma^{s_i}_{G,\mathbf{k}}+\int_{L^{-1}}\frac{d^3p_1}{(2\pi)^3}\frac{d^3p_2}{(2\pi)^3}(2\pi)^3\delta^3(\mathbf{p}_1+\mathbf{p}_2-\mathbf{k}) F(p_1,p_2,k)(\gamma^{s_i}_{G,\mathbf{p_1}}\chi_{G,\mathbf{p_2}}+\gamma^{s_i}_{G,\mathbf{p_2}}\chi_{G,\mathbf{p_1}})+...,
\label{nonlocalzeta}
\ee
where we take
\be
F(k_1,k_2,k_3)=f_{\gamma\gamma\chi}^{\rm eff}(k_p)\sum_ja_j\left(\frac{k_1}{k_p}\right)^{m_{1,j}}\left(\frac{k_2}{k_p}\right)^{m_{2,j}}\left(\frac{k_3}{k_p}\right)^{m_{3,j}}\;.
\label{kernel}
\ee
For long wavelength modes of the $\chi$ field, the tensor power spectrum is shifted by
\bea
\label{Pgammaobs}
P\o_{\gamma}&=&P_{\gamma}\left[1+\Delta\chi(k)\right]^2\\\nonumber
\Delta\chi(k)&=&a\,f_{\gamma\gamma\chi}^{\rm eff}(k_p)\left[\left(\frac{k}{k_p}\right)^{m_1+m_3}\chi_{Gl}^{(m_2)}+\left(\frac{k}{k_p}\right)^{m_2+m_3}\chi_{Gl}^{(m_1)}\right]\\\nonumber
\chi_{Gl}^{(m_L)}&\equiv&\int_{L^{-1}}^{M^{-1}}\frac{d^3p}{(2\pi)^3}\chi_{\mathbf{p}}\left(\frac{p}{k_p}\right)^{m_L}\;.
\eea
With this parameterization, long wavelength modes of the $\chi$ field can shift the locally observed tilt of the tensor power spectrum. In the case that the first term in $\Delta\chi$ dominates, we can again define $m_S=m_1+m_3$, $m_L=m_2$ and then the shift is approximately
\begin{equation}
\label{eq:Deltant}
\Delta n_{t}(k)  \approx  2\Delta\chi(k)\,m_S \;.
\end{equation}
The quantity $m_S$ is zero for an exactly scale-invariant, local type model and more generally cannot be too large if we want to require weak non-Gaussianity for all fields. Depending on the coupling of $\chi$ to the scalar curvature, this physics may also introduce a shift in the locally observed scalar power spectrum, the tensor-to-scalar ratio, and a `fossil' signature in the off-diagonal part of the scalar power spectrum \cite{Jeong:2012df}, which would be an interesting complementary observable. From these expressions, it looks possible to find scenarios where the locally observed tensor power spectrum would be shifted from red to blue and vice-versa, but a full analysis along the lines of Section \ref{sec:ObsConsequences} should be performed to check consistency with all observables. 

\section{Discussion and conclusions}
\label{sec:conclude}
Non-Gaussianity that couples the statistics of fluctuations on observable scales to wavemodes spanning super-Hubble scales can bias cosmological statistics measured by an observer in a local Hubble volume. Previous work showed that the relative amplitudes of the power spectrum and non-Gaussianity ($f^{\rm local}_{\rm NL}$) can vary in observable subvolumes. In this work we have shown that the spectral index can also vary by enough to be interesting, $|\Delta n_s|\simeq0.04$. The scaling of the squeezed limit of the bispectrum can also be shifted, which is relevant for constraints on non-Gaussianity from galaxy bias. These results show that in spite of the excellent precision of the measurements from the {\it Planck} satellite (especially $n\o_s=0.9603\pm0.0073$ \cite{Ade:2013uln} and constraints on non-Gaussianity), the door is open for a significant cosmic variance uncertainty in comparing our observed patch of the universe to any particular inflation theory - even leaving aside issues with eternal inflation. Moreover, rather than just presenting a new source of uncertainty from the super-Hubble background, the correlation between bispectral running in a super-Hubble volume and subvolume power spectrum measurements reopens the door for inflationary models with flat or bluer super-Hubble spectral indices, $n_s \neq 0.96$, provided they also have scale-dependent local non-Gaussianity. This may be particularly useful for hybrid inflation. 

The numbers measured by the {\it Planck} satellite are consistent with a range of levels of non-Gaussianity in a post-inflationary volume, given a model for the statistics in that volume. For example, we recover the observed power spectrum and spectral index, and satisfy current constraints on $f_{\rm NL}^{\rm CMB}$ for a post-inflationary volume with
\begin{itemize}
\item  No local type non-Gaussianity, an arbitrary number of extra e-folds, and any behavior of the power spectrum on superhorizon scales.
\item Constant $f^{\rm local}_{\rm NL}=5$. We observe $f_{\rm NL}^{\rm local}=8$ if, for example, the spectral index is a constant $n_s=0.96$ over about 200 extra e-folds of inflation and our Hubble patch sits on top of a 2-sigma under density.
\item Local non-Gaussianity with constant $f^{\rm local}_{\rm NL}=15$. We observe $f_{\rm NL}^{\rm local}=11$ if, for example, the spectral index is a constant $n_s=0.96$ over about 150 extra e-folds of inflation and our Hubble patch sits on top of a 2-sigma over-density.
\item Scale-dependent non-Gaussianity with $f_{\rm NL}(k_p)=-2$, $n_f=0.04$, $n_\zeta=0.93$, and $n_s=0.935$. We would observe $f_{\rm NL}(k_p)=-1$ and $n_s\o=0.956$ if our Hubble patch sits on top of a 2-sigma under density in a volume with about 190 extra e-folds.
\item Scale-dependent non-Gaussianity with $f_{\rm NL}(k_p)=20$, $n_f=0.03$, $n_\zeta=0.95$, and $n_s=1.005$. We would observe $f_{\rm NL}(k_p)=2.5$ and $n_s\o=0.975$ if our Hubble patch sits on top of a 0.2-sigma over density in a volume with about 280 extra e-folds.
\end{itemize}
In contrast, we could design an inflation model to have parameters roughly consistent with {\it Planck} data, say $f_{\rm NL}(k_p)=5$, $n_f=0.1$, $n_\zeta=0.98$, and $n_s=0.982$. However, if the model allows about 400 extra e-folds of inflation, and our Hubble patch were to sit on a 2-sigma over density, we would observe $f_{\rm NL}(k_p)=4$ and $n_s\o=1.013$.

These results demonstrate that predictions for our observations in any scenarios with local type non-Gaussianity must be given statistically. To turn the picture around, they also suggest a new route to understanding whether observations can give us any hints about the size of the universe beyond what is directly observable. Previous ideas focused on topologically finite universes (also significantly constrained by {\it Planck} \cite{Ade:2013vbw}) or on evidence for or against a nonperturbatively connected multiverse from bubble collisions \cite{Chang:2008gj, Feeney:2012hj, Osborne:2013hea} or curvature \cite{Kleban:2012ph, Guth:2012ww}. While observations will probably never tell us how long inflation lasted, our work suggests they may at least tell us if that uncertainty is relevant to our interpretation of the data we do have. 

From a cosmic variance point of view, we are fortunate that there is so far no detection of local type non-Gaussianity. We have shown that future observations could push the mode-coupling uncertainties we have considered here into irrelevance\footnote{Or, at least, back into the realm of eternal inflation; see the discussion about perturbative fluctuations below Eq.~(\ref{eq:zetaagain}).} if primordial local non-Gaussianity can be constrained to be $|f_{\rm NL}| < 1$. Even if $|f_{\rm NL}|>1$ is observed, tests for the running of the spectral index, any scale-dependence of $|f_{\rm NL}|$,  and any evidence for extra fields through isocurvature modes or `fossil' relics hiding in the off-diagonal power spectrum could still limit the size of any subsampling uncertainty. For example, if a blue tilt to $f_{\rm NL}$ is ruled out, biasing of the spectral index is unlikely for single-source models with $n_{\zeta}$ and $n_f$ constant on all scales. Of course, making these observations statistically well-defined depends on comparing particular competing models. It would be particularly interesting if those models had other cosmological implications related to the size of the universe\footnote{We thank S. Dubovsky for bringing this reference to our attention.} \cite{Kaplan:2008ss}. It would also be worthwhile to investigate the generic behavior of the local ansatz beyond $f_{\rm NL}$ alone with scale-dependent coefficients, along the lines of the analysis in \cite{Nelson:2012sb}. It may be that there are statistically natural values for the spectral index in typical small subvolumes. Then, stronger conclusions about generic cosmic variance of the spectral index might be possible.  However, it is already clear that if improved limits on the amplitude and scale-dependence of non-Gaussianity can be reached, we could close the window of observational access to a perturbatively connected larger universe.\\

{\bf Acknowledgements}

We thank Chris Byrnes, Bhaskar Dutta, Louis Leblond and Marilena LoVerde for useful suggestions and discussions about this work. The work of J.~B.~is supported in part by Department of Energy grant DE-FG02-04ER41291. The work of J.~K.~ is supported in part by Department of Energy grants DE-FG02-04ER41291 and DE-FG02-13ER41913.  The work of S.~S.~is supported in part by the National Aeronautics and Space Administration under Grant No. NNX12AC99G issued through the Astrophysics Theory Program. In addition, S.~S.~ thanks the organizers of the Primordial Cosmology Program at KITP for hospitality while this work was being completed and for support by the National Science Foundation under Grant No. NSF PHY11-25915.  J.~K.~thanks the Center for Theoretical
Underground Physics and Related Areas (CETUP* 2013) in South Dakota
for its support and hospitality while this work was being completed.
E.~N.~is supported by the Eberly Research Funds of The Pennsylvania State University. The Institute for Gravitation and the Cosmos is supported by the Eberly College of Science and the Office of the Senior Vice President for Research at the Pennsylvania State University.

\end{document}